\documentclass[a4paper,twocolumn,10pt,accepted=2022-08-31]{quantumarticle}
\pdfoutput=1

\usepackage[utf8]{inputenc}
\usepackage[english]{babel}
\usepackage[T1]{fontenc}
\usepackage[numbers,sort&compress]{natbib}
\usepackage{listings}

\usepackage{amsmath}
\usepackage{amssymb}

\usepackage{lipsum}
\usepackage{placeins}
\usepackage{braket}
\usepackage{pgfplots}
\usepackage{mathrsfs}
\usepackage{physics}
\usepackage{bbold}
\usepackage[normalem]{ulem}
\usepackage{comment}
\usepackage{hhline}

\newcommand\blfootnote[1]{%
  \begingroup
  \renewcommand\thefootnote{}\footnote{#1}%
  \addtocounter{footnote}{-1}%
  \endgroup
}

\newcommand{\beq}{\begin{equation}}
\newcommand{\eeq}{\end{equation}}

\def\la{{\langle}}
\def\ra{{\rangle}}

\newcommand{\ie}{\textit{i.e.} }
\newcommand{\eg}{\textit{e.g.} }

\pgfplotsset{compat=1.17}
\begin{document}

\title{Algebraic Bethe Circuits}

\author{Alejandro Sopena*}
\affiliation{Instituto de F\'{\i}sica Te\'{o}rica, UAM/CSIC, Universidad Aut\'{o}noma de Madrid, Madrid, Spain}
\author{Max Hunter Gordon*}
\affiliation{Instituto de F\'{\i}sica Te\'{o}rica, UAM/CSIC, Universidad Aut\'{o}noma de Madrid, Madrid, Spain}
\affiliation{Theoretical Division, Los Alamos National Laboratory, Los Alamos, NM 87545, USA}
\author{Diego García-Martín}
\affiliation{Barcelona Supercomputing Center, Barcelona, Spain}
\affiliation{Instituto de F\'{\i}sica Te\'{o}rica, UAM/CSIC, Universidad Aut\'{o}noma de Madrid, Madrid, Spain}
\author{Germ\'{a}n Sierra}
\affiliation{Instituto de F\'{\i}sica Te\'{o}rica, UAM/CSIC, Universidad Aut\'{o}noma de Madrid, Madrid, Spain}
\author{Esperanza L\'{o}pez}
\affiliation{Instituto de F\'{\i}sica Te\'{o}rica, UAM/CSIC, Universidad Aut\'{o}noma de Madrid, Madrid, Spain}
\blfootnote{* The first two authors contributed equally.}
\begin{abstract}
The Algebraic Bethe Ansatz (ABA) is a highly successful analytical method used to exactly solve several physical models in both statistical mechanics and condensed-matter physics. Here we bring the ABA into unitary form, for its direct implementation on a quantum computer. This is achieved by distilling the non-unitary $R$ matrices that make up the ABA into unitaries using the QR decomposition. Our algorithm is deterministic and works for both real and complex roots of the Bethe equations. We illustrate our method on the spin-$\frac{1}{2}$ XX and XXZ models.  We show that using this approach one can efficiently prepare eigenstates of the XX model on a quantum computer with quantum resources that match previous state-of-the-art approaches. We run small-scale error-mitigated implementations on the IBM quantum computers, including the preparation of the ground state for the XX and XXZ  models on $4$ sites. Finally, we derive a new form of the Yang-Baxter equation using unitary matrices, and also verify it on a quantum computer.
\end{abstract}
\maketitle

\begin{figure}[ht!]
    \centering
    \includegraphics[width = \columnwidth]{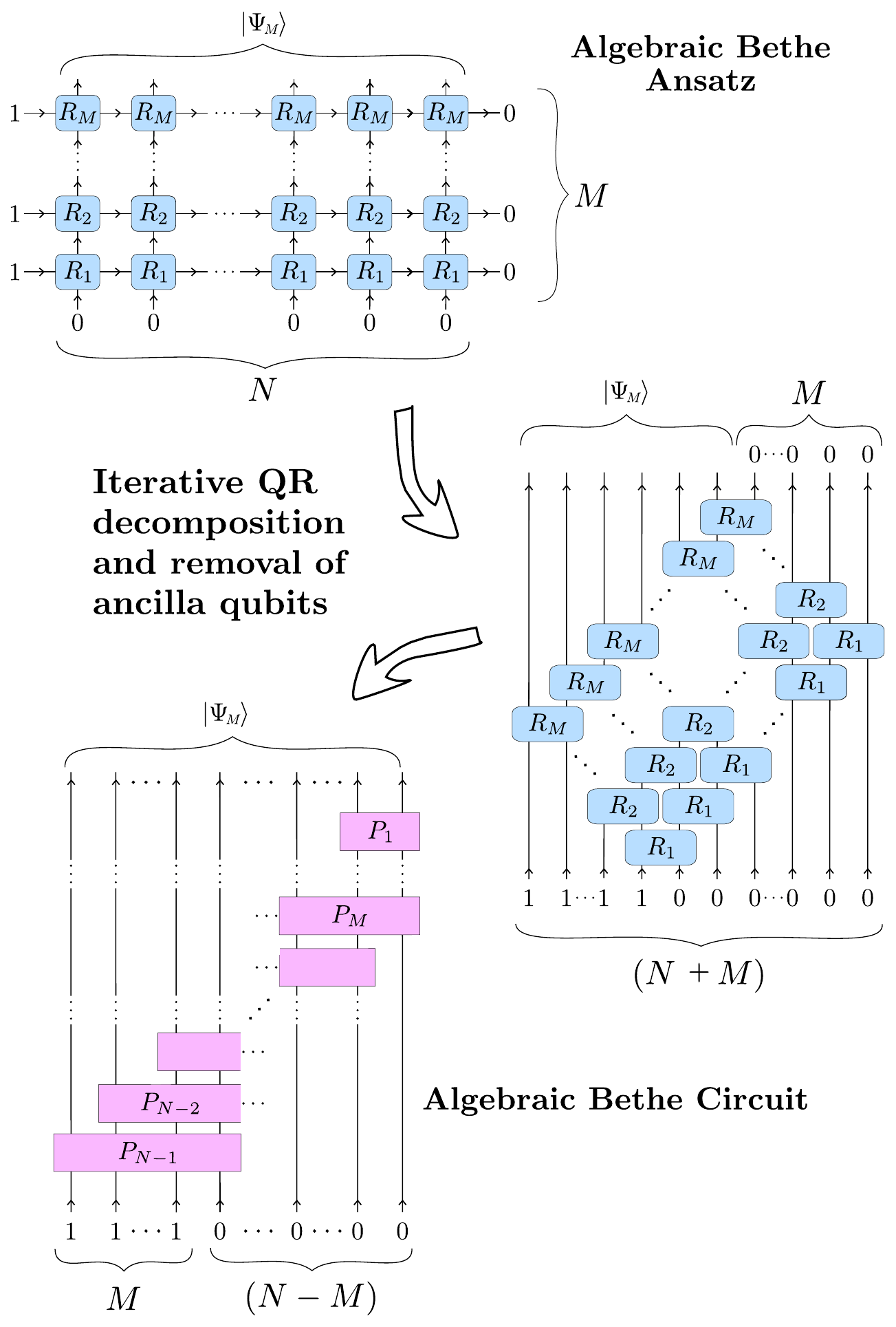}
    \caption{A Bethe ansatz eigenstate $\ket{\Psi_M}$ for $M$ magnons on $N$ sites, converted into a quantum circuit of $N$ qubits by computing iterative QR decompositions and removing the ancilla qubits, as explained in the main text. The $R_{j}$ are the tensors defined in~\eqref{eqn:R} that satisfy the YB equation, and the $P_{k}$ are unitary matrices of dimension $2^{n+1}\times 2^{n+1}$ with $n=\min(k,M)$. These matrices need to be compiled to quantum gates before implementation on a quantum computer.}
    \label{fig:ABA_MPS_conv}
\end{figure}

\section{Introduction}

One of the most widely-recognized applications of quantum computing is the efficient simulation of many-body quantum systems~\cite{Lloyd1073}. Simulating such systems using classical devices generally requires computational resources that scale exponentially with the size of the system. Quantum computers on the other hand are naturally suited to this task, being themselves quantum systems and hence overcoming the classical exponential scaling.
In this context, preparing Bethe Ansatz (BA) eigenstates on a quantum computer is attracting increasing attention~\cite{nepomechie2021bethe,vandyke2021preparing,vandyke2021preparingopen, li2022bethe}.

The BA is an extremely successful classical method for exactly solving one-dimensional (1D) quantum models, \eg the Heisenberg, Hubbard or Kondo models~\cite{Bethe1931, Korepin1993kvr, faddeev1996algebraic, gomez_quantum_1996}. It reduces the difficult problem of diagonalizing the Hamiltonian to finding the solutions of a set of algebraic equations. In many cases, it is possible to numerically solve these equations, which in turn allows one to calculate the eigenvalues and the eigenvectors of the system of interest. These quantities are computed differently depending on whether the coordinate Bethe Ansatz (CBA) or the Algebraic Bethe Ansatz (ABA) is used~\cite{gomez_quantum_1996}. In both cases, the eigenstates are represented as complex mathematical expressions. As a result, this method does not allow direct access to some physical quantities, such as high-order and long-range correlation functions, which have proved challenging to compute both analytically and numerically~\citep[and references therein]{Balazs2017}.
This motivates the construction of such states directly on a quantum computer. Once the eigenstates are experimentally available, all the correlation functions can be readily computed from measurements. Furthermore, these states can be used to initialise other quantum algorithms or benchmark quantum hardware.

\FloatBarrier

To this end, a quantum algorithm for the preparation of the eigenstates of the spin-$\frac{1}{2}$ XXZ model in 1D was introduced in Refs.~\cite{vandyke2021preparing,vandyke2021preparingopen}, based on the CBA. This algorithm is probabilistic and works for real-valued solutions of the Bethe equations. Its circuit depth is polynomial in both the system size and the number of magnons, or down spins. However, it was recently shown~\cite{li2022bethe} that the success probability of obtaining the desired eigenstate upon measurement of the ancillary qubits decreases super-exponentially with the number of magnons for large chains.
The use of a variational approach for the preparation of integrable-system eigenstates, which does not rely on knowledge coming from the BA, has also been considered~\cite{ho2019efficient,BravoPrieto2020scalingof, wieresma2020exploring,jattana2022assessment}. Variational quantum algorithms are known to suffer from exponentially-vanishing gradients~\cite{McClean_2018,cerezo2020variationalreview}. This problem is certain to appear when the number of magnons scales with the system size~\cite{larocca2021diagnosing}.
Moreover, even when the number of magnons is fixed this issue is likely to persist for large chains due to the effect of noise~\cite{2021NIBP}. 

Another approach could be to prepare eigenstates of interest using Matrix Product State (MPS) optimization~\cite{schollwock2005density} and then use further optimization techniques to map the classically tractable state to a quantum circuit. However, calculating an accurate classical representation depends on the structure of entanglement of the systems and will not be trivial in all phases or the entire spectrum of states~\cite{Pirvu_2012}. Furthermore, once a classical representation of the state of interest is calculated constructing a faithful quantum circuit to represent it in a quantum computer may also be difficult and is an active field of research~\cite{dborin2022matrix,smith2022crossing}.

In this paper, we present a quantum algorithm based on the ABA for the preparation of BA eigenstates. Contrary to~\cite{vandyke2021preparing}, it also works for complex solutions of the Bethe equations. 
The main difficulty encountered when directly trying to convert the ABA into a quantum circuit is that the matrices $R$ are not unitary. Besides, the translation of the ABA into a circuit requires the use of ancillas which need to be projected onto the $|0\dots0\ra$ state at the end of the computation.
This would yield a probabilistic algorithm should a direct translation be employed. We circumvent both of these issues by iteratively computing the QR decomposition~\cite{golub_matrix_1996} of the non-unitary matrices appearing in the ABA. This then allows us to obtain the desired eigenstate as the output of a quantum circuit with no ancillary qubits, which we refer to as an Algebraic Bethe Circuit (ABC).
The process is sketched in Fig.~\ref{fig:ABA_MPS_conv}. We note that
quantum circuits based on unitary $R$ matrices were explored to compute infinite temperature correlation functions  in~\cite{claeys2021correlations}. 

We focus on the paradigmatic 1D anti-ferromagnetic spin-$\frac{1}{2}$ XXZ model with periodic boundary conditions for the concrete analysis. The complexity of calculating the ABC unitaries scales linearly in the number of sites (qubits), but exponentially in general with the number
of magnons. There is the additional complexity of compiling the calculated unitaries, which may also scale exponentially. This means that our algorithm is in general only applicable to a small number of magnons. In spite of that, it could allow for the preparation of states on near-term quantum hardware that have been challenging before.
Besides, we find that its application on the XX model is efficient. This model describes free spinless fermions via a Jordan-Wigner transformation. Here, the ABCs match the performance of state-of-the-art algorithms for the preparation of fermionic states~\cite{2018Babush,2018Boixo, arute2020observation}, both in the number of gates necessary and the circuit depth.

From a theoretical standpoint, the ABCs offer an alternative approach towards finding exact circuits for integrable quantum many-body systems~\cite{Verstraete_2009, CerveraLierta2018exactisingmodel}.
Along these lines, we derive a novel version of the Yang-Baxter equation in terms of unitary matrices which can be tested on quantum hardware.
Interestingly, the BA can be interpreted as a MPS ~\cite{2003_bethe_MPS,KatsuraMPS,Murg_2012}.
Our algorithm for the distillation of unitaries from the ABA appears closely related to the transformation of an MPS into canonical form~\cite{perezgarcia2007matrix}, which in turn has a direct translation to a quantum circuit.
Hence our method to obtain ABCs should also prove relevant for the circuit implementation of general MPS with low bond dimension. Let us mention that several works have considered the implementation of tensor-network states on a quantum computer~\cite{ShiJu2020,smith2022crossing,Lin_compressed,Barratt_2021,2021_holographic_QC,Lin_compressed,hagh2021variational}.

All current quantum computers suffer from significant hardware noise. Therefore, error mitigation will play an essential role in all near-term quantum algorithms and simulations. As such, we explore the performance of several state-of-the-art error mitigation methods on mitigating observables produced by our algorithm.

The paper is structured as follows. In Section~\ref{sec:ABA} we briefly introduce the ABA. Section~\ref{sec:ABC} contains a detailed explanation of its transformation into a quantum circuit, together with a preliminary discussion on the decomposition of the ABC unitaries in terms of elementary quantum gates. The unitary version of the YB equation is discussed in Section~\ref{sec:unitary_YB}. Section~\ref{sec:experiments} contains small-scale error-mitigated implementations of the ABCs on quantum hardware. In particular, we prepare plane wave states on 8 sites and implement the ground state of the XX and XXZ models on 4 sites using the IBM cloud quantum computers. Section~\ref{sec:experiments} also includes a test of the YB equation on quantum hardware.
Finally, we conclude in Section~\ref{sec:discussion} with a discussion of our results. Several technical details are confined to Appendices.

\section{The Algebraic Bethe Ansatz}
\label{sec:ABA}
The Algebraic Bethe Ansatz~\cite{Korepin1993kvr,faddeev1996algebraic,gomez_quantum_1996} is a powerful classical technique to solve one-dimensional quantum integrable vertex models. These models fulfill the so-called Yang-Baxter (YB) equation~\cite{baxter_exactly_1985}, and are characterised by an extensive set of conserved quantities. They have been greatly studied over the last century in the context of quantum many-body physics~\cite{sutherland_beautiful_2004,mussardo_statistical_2020}.

The ABA can be used to calculate the exact eigen-spectrum of a large class of Hamiltonians. Among these models, a prominent example is the 1D spin-$\frac{1}{2}$ anti-ferromagnetic XXZ Hamiltonian on $N$ sites with periodic boundary conditions,
\begin{equation}
H_{\textrm{XXZ}} = \sum_{j=1}^N  \left( \sigma^X_j \sigma^X_{j+1} +  \sigma^Y_j \sigma^Y_{j+1} + \Delta \,  \sigma^Z_j \sigma^Z_{j+1} \right)\,,
\label{eqn:hamil}
\end{equation}
where $\{\sigma^X_j,\sigma^Y_j,\sigma^Z_j\}$ are the Pauli matrices acting on the $j$-th spin, and  $\sigma_{N+1}\equiv \sigma_1$. The parameter $\Delta$ introduces an anisotropy in the chain.
When $\Delta=1$ we recover the isotropic anti-ferromagnetic Heisenberg spin chain or XXX model, which has $SU(2)$ symmetry.
For other values of $\Delta$, only a $U(1)$ symmetry is present.
A set of algebraic equations needs to be solved in order to construct the eigenstates of the Hamiltonian. These are the celebrated Bethe equations, whose solutions or roots are referred to as rapidities which we denote as $\{\lambda_j\}_{j=1}^M$. The number of Bethe roots $M$ describes the number of magnons or spin-down waves composing the state.

We shall consider the XXZ model with anisotropy in the range $\Delta \in (-1, 1]$. The corresponding Bethe equations are given by   
\begin{equation}
    \left(\frac{\sinh{\left(\gamma\frac{\lambda_j+i}{2}\right)}}{\sinh{\left(\gamma\frac{\lambda_j - i}{2}\right)}}\right)^{N}=\prod_{\substack{k=1 \\ k \neq j}}^{M} \frac{\sinh{\left(\gamma\frac{\lambda_j-\lambda_k+2i}{2}\right)}}{\sinh{\left(\gamma\frac{\lambda_j-\lambda_k-2i}{2}\right)}}\ ,
\label{eqn:Bethe_eqs}
\end{equation}
where $\cos{\gamma}=\Delta$, with $\gamma\in[0,\pi)$. The ground state of a chain with an even number of sites is built out of $M=N/2$ magnons whose rapidities are real. In general, excited eigenstates however contain complex rapidities, that must come in conjugate pairs to guarantee that the energy is real-valued. The Bethe equations are typically solved by numerical methods~\cite{giamarchi_quantum_2003}.

The BA eigenstates can be represented as a product of operators acting on the vacuum state $\ket{vac}$ as
\begin{equation}
    \ket{\Psi_M} = B(\lambda_{M})\cdots B(\lambda_{2})B(\lambda_{1})\ket{vac}\,.
    \label{eqn:BA_state}
\end{equation}
In the XXZ model, the $\ket{vac}$ state is the product state with all spins up, \ie $|0\dots0\ra$, and $B(\lambda)$ is an operator that creates one magnon. This operator can be represented as the contraction of a network of four-index tensors (see Appendix~\ref{app:integrable} for a more detailed explanation)
\begin{equation}
       \vcenter{\hbox{\includegraphics[width=0.85\columnwidth]{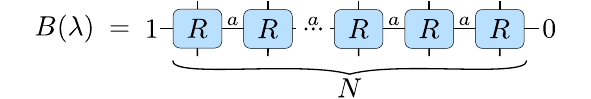}}} \ .
       \label{B}
\end{equation} 
These tensors, reshaped into $4\times 4$ matrices, are the $R$ matrices, which depend on the rapidities and satisfy the YB equation.
Each $R(\lambda)$ gives a map $\mathcal{H}_a \otimes \mathcal{H}_j \rightarrow \mathcal{H}_j \otimes \mathcal{H}_a$, where $\mathcal{H}_j$ is the Hilbert space of the $j$-th spin and $\mathcal{H}_a$ is an auxiliary Hilbert space of dimension 2. The auxiliary space is shared by all the matrices involved in the creation of one magnon, \ie sharing a common rapidity $\lambda$.

We have used the following convention
\begin{equation}
       \vcenter{\hbox{\includegraphics[width=0.85\columnwidth]{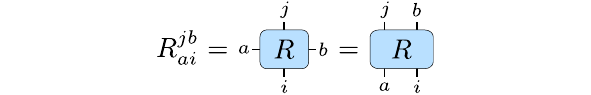}}} \ ,
       \label{eq:R_index}
\end{equation}
with the last equality making clear the matrix interpretation of the tensors appearing in~\eqref{B}. The input and output indices $ai$ and $jb$ correspond respectively to the columns and rows of the $R$ matrix.

The $R$ matrix of the XXZ model is
\begin{equation}
R = \rho\left( \begin{array}{cccc}
1 & 0 & 0 & 0 \\
0 & s_1 & s_2 & 0 \\
0 & s_2 & s_1 & 0 \\
0 & 0 & 0 & 1 \\
\end{array}
\right)\,,
\label{eqn:R}
\end{equation}
with $\rho$ a complex number and the parameters $s_1,s_2$ satisfying
\begin{equation}
1+s_2^2-s_1^2= 2s_2\Delta \ .
\label{eqn:Delta}
\end{equation}
In terms of the rapidity, this equation is solved by
\begin{equation}
 s_1(\lambda)=\frac{\sinh{i\gamma }}{\sinh{\left(\gamma\frac{\lambda+i}{2}\right)}} \quad,\quad s_2(\lambda) = \frac{\sinh{\left(\gamma\frac{\lambda - i}{2}\right)}}{\sinh{\left(\gamma\frac{\lambda+i}{2}\right)}} \ .
 \label{eqn:R_param}
 \end{equation}
The second parameter has a direct physical interpretation as the magnon quasi-momentum, $s_2=e^{i p}$. It is important to note that the $R$ matrix is excitation preserving as a consequence of the $U(1)$ symmetry of the Hamiltonian~\eqref{eqn:hamil}.

\section{From ABA to ABC}
\label{sec:ABC}
\subsection{Detailed method}
Our aim is to construct quantum circuits based on the ABA, for the direct preparation of eigenstates of integrable vertex models on quantum hardware.
We will call these circuits Algebraic Bethe Circuits (ABCs).

An ABA eigenstate for $M$ magnons on $N$ sites is shown in Fig.~\ref{fig:ABA_MPS_conv}, where it is transformed into a suggestive form with the apparent structure of a quantum circuit. The basic cell $\mathscr{R}_T$ that repeats itself throughout this circuit is
\begin{equation}
    \vcenter{\hbox{\includegraphics[width=0.85\columnwidth]{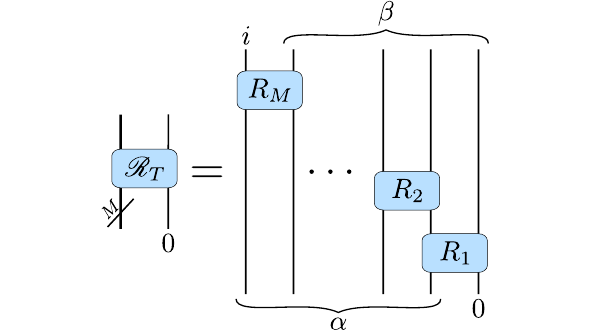}}} \ ,
\label{calR}
\end{equation}
where $R_j\equiv R(\lambda_j)$. The problem encountered when directly trying to transform this cell into a quantum gate is that the matrices $R$ are in general not unitary (see Appendix~\ref{app:R_non_unitary}). 

To stress the crucial difference between unitary and non-unitary matrices, we shall use rounded-corner rectangles for the latter. 
The complete ABA network can be recast in terms of the $\mathscr{R}_T$ cells as
\begin{equation}
    \vcenter{\hbox{\includegraphics[width=0.85\columnwidth]{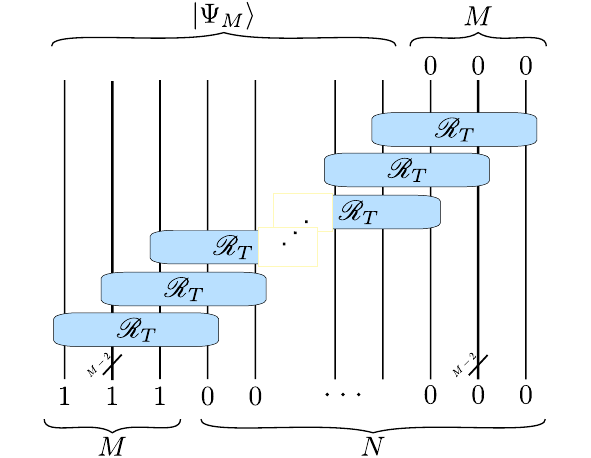}}} \ .
    \label{RTcircuit}
\end{equation}
The $M$ rightmost qubits, both at the input and output, are in the fixed state $\ket{0}$. 
They can be considered ancillary qubits. Keeping them in the final quantum circuit we are seeking would result in a probabilistic algorithm. A severe reduction of the success probability with increasing $M$ is to be expected due to the curse of dimensionality. The probabilistic quantum algorithm based on the CBA proposed in~\cite{vandyke2021preparing} was recently shown to suffer from a similar problem~\cite{li2022bethe}. 

Both the conversion of~\eqref{RTcircuit} into a quantum circuit and the removal of the ancillary qubits can be addressed by utilising the QR decomposition as our main tool.
To be more precise, any $m \times n$ matrix can be written as the product of two matrices $Q \cdot R$. When $m \leq n$, $Q$ is a $m \times m$ unitary and $R$ an $m \times n$ rectangular matrix with vanishing entries below the main diagonal. When $m > n$, $Q$ is a $m \times n$ isometry ({\it i.e.} $Q^\dag Q= \mathbb{1}_n$) and $R$ an $n \times n$ upper triangular matrix. We note that the QR decomposition was previously used in the derivation of efficient quantum circuits in~\cite{troyer_fermions,2018Boixo, 2018Babush, arute2020observation}.

We start our protocol at the top-rightmost basic cell. All the information in this cell can be encoded in a $2 \times 2^M$ matrix defined by
\begin{equation}
    \vcenter{\hbox{\includegraphics[width=0.85\columnwidth]{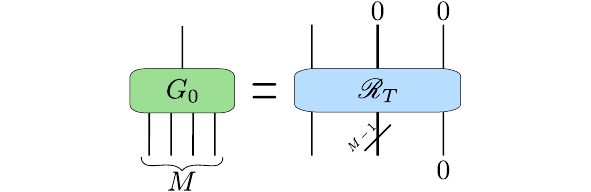}}} \ .
    \label{G0}
\end{equation}
Working directly with $G_0$ allows us to eliminate the rightmost ancillary qubit. The matrix $G_0$ can then be absorbed into the second ${\mathscr R}_T$ cell. This defines a $4 \times 2^M$ matrix which renders the second rightmost qubit unnecessary. We apply now the QR decomposition to this matrix, obtaining
\begin{equation}
    \vcenter{\hbox{\includegraphics[width=0.85\columnwidth]{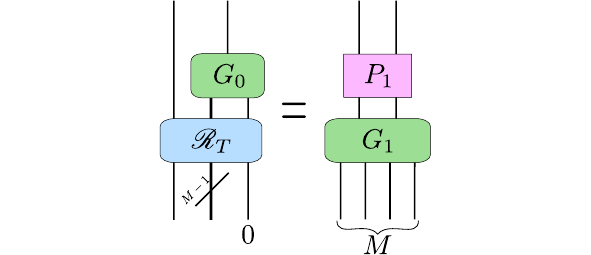}}} \ .
    \label{P1G_1}
\end{equation}
We have distilled the first gate of the deterministic quantum circuit for the construction of Bethe eigenstates, the two-qubit unitary $P_1$. The non-unitary remainder $G_1$  is then absorbed into the next basic cell and the QR decomposition is computed again for the new non-unitary matrix. As before, one more ancillary qubit is eliminated. This process is iterated. At each step a unitary gate $P_k$ acting on $k+1$ qubits is obtained. After $M-1$ iterations all ancillary qubits have been removed and the $M$ top-right basic cells have been substituted by the circuit
\begin{equation}
    \vcenter{\hbox{\includegraphics[width=0.85\columnwidth]{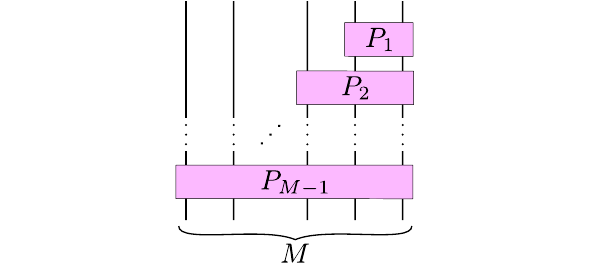}}} \ .
\end{equation}

For $k \geq M$  
each new step is described by the recursion relation
\begin{equation}
    \big( \mathbb{1} \otimes G_{k\!-\!1} \big) {\mathscr R}_T \ket{0}= \big( P_k \ket{0} \big)  G_k \ ,
\end{equation}
or equivalently
\begin{equation}
    \vcenter{\hbox{\includegraphics[width=0.85\columnwidth]{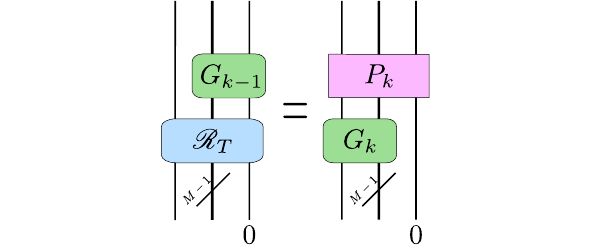}}} \ .
    \label{rec}
\end{equation}
The LHS defines a $2^{M\!+\!1} \times 2^M$ matrix. 
The matrix $Q$ resulting from its QR decomposition is in this case an isometry, which determines $P_k \ket{0}$. This information can be completed at our best convenience to define the $M+1$ qubit gate $P_k$ that is to be implemented on the quantum circuit. 

In order to solve~\eqref{rec}, we multiply both sides with the Hermitian conjugate to obtain a recursion relation involving only the upper triangular matrices $G_k$, \ie
\begin{equation}
    \bra{0} {\mathscr R}_T^\dag \, \big( \mathbb{1} \otimes 
    G_{k\!-\!1}^\dag G_{k\!-\!1} \big) {\mathscr R}_T \ket{0} = G_k^\dag G_k \ .
    \label{rec2}
\end{equation}
Once the solution $G_{k}$ has been calculated, we extract the $P_k$ from
\begin{equation}
     P_{k} \ket{0} = \big(\mathbb{1}\otimes G_{k-1}\big) \big( \mathscr{R}_{T} \ket{0} \! \big) G_{k}^{-1} \ .
     \label{PfromG}
\end{equation}
For $k<M$ a similar relation holds
\begin{equation}
    P_k =\big(\mathbb{1}\otimes G_{k-1}\big) \big( \mathscr{R}_{T} \ket{0} \! \big) H_{k} \ .
    \label{PfromG2}
\end{equation}
Recall that $G_{k<M\!-\!1}$ are not square matrices and hence do not have an inverse. However, there exists a non-square upper triangular matrix $H_k$ satisfying $G_k H_k=\mathbb{1}$, which is thus sufficient for our purposes.
Although the dimensions of $G_k$ vary in the first iterations, the product $G_k^\dag G_k$ always defines a $2^M \times 2^M$ matrix. This renders the previous equation applicable to all iterations.

The protocol ends at the bottom-left cell
\begin{equation}
    \vcenter{\hbox{\includegraphics[width=0.85\columnwidth]{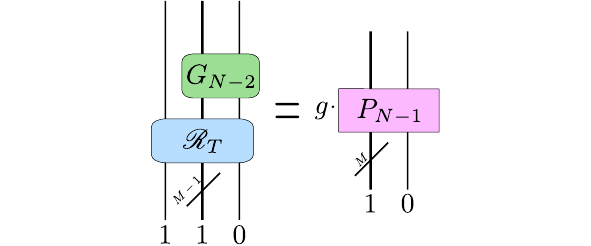}}} \ .
    \label{eq:end_method}
\end{equation}
The correct normalization of the output Bethe state $|\Psi_M \rangle$ from the ABA requires one to suitably adjust the global factor $\rho$ of the $R$ matrix~\eqref{eqn:R}. If this is done, $g$ will be a trivial phase.
Alternatively, we might ignore the difficult issue of finding the appropriate $\rho$ for the desired eigenstate and just set $\rho=1$.
The output state from~\eqref{RTcircuit}
will not be properly normalized, but the problem is simply solved by discarding the global factor $g$. We adopt this convention in the following. The previous relation~\eqref{eq:end_method} only determines the action of $P_{N-1}$ on the state $\ket{1\dots 10}$.
This last unitary can be otherwise chosen freely.

We have thus obtained a deterministic quantum circuit for the preparation of BA eigenstates on $N$ sites requiring only $N$ qubits.
The upper triangular structure of $G_k$ is key to the feasibility of our protocol. It implies the reduction of~\eqref{rec2} to a nested system of equations. Substituting the solution of one equation into the next, we only need to care for one entry of $G_k$ at a time. Moreover the inverse of a triangular matrix, required to obtain the unitaries $P_k$, is not expensive to calculate.
The bottleneck of the procedure is the exponential growth in the number of equations with the number of magnons. Nevertheless, we have found that this is not prohibitive to explore cases of interest. 

Two important comments on the QR decomposition should be added. The matrices $Q$ and $R$ are not fully determined but enjoy the gauge freedom
\begin{equation}
    Q \rightarrow Q \, D^{-1}\ , \hspace{5mm} R \rightarrow D \, R \ ,
    \label{diagonals}
\end{equation}
where $D$ is an arbitrary diagonal matrix containing complex phases. This freedom will be crucial below, when compiling the unitaries to elementary quantum gates. Additionally, the QR decomposition is compatible with the $U(1)$ symmetry of the XXZ model. Hence, all matrices $G_k$ and $P_k$ will conserve individually the number of excitations.

We end with a final remark. It has been shown that the BA can be interpreted in the language of tensor networks as a Matrix Product State (MPS)~\cite{2003_bethe_MPS,KatsuraMPS,Murg_2012}.
Our algorithm for the distillation of unitaries from the ${\mathscr R}_T$ matrices is closely related to the transformation of an MPS into canonical form~\cite{perezgarcia2007matrix}. The tensors of a canonical MPS define isometries, and thus have a direct translation to a quantum circuit. The main difference with our approach lies in the elimination of the ancillary qubits, which renders the circuit deterministic. Hence the ABCs
could prove relevant, not only for the preparation of eigenstates of integrable spin chains, but also for general MPS of low bond dimension.
We note that the
implementation of tensor-network states on a quantum computer has been addressed in~\cite{ShiJu2020,smith2022crossing,Lin_compressed,Barratt_2021,2021_holographic_QC,Lin_compressed,hagh2021variational}.

\subsection{One-magnon solution}
We illustrate our method by explicitly deriving the circuit that constructs one-magnon states for the XXZ model.
In order to study the properties of the $P_k$ gates, we find it convenient to use  the magnon quasi-momentum $p$ instead of the rapidity $\lambda$ as the basic variable. Indeed, the $R$ matrix~\eqref{eqn:R} has a simple expression in terms of $s_1$ and $s_2=e^{i p}$. Besides, the integrability constraint~\eqref{eqn:Delta} turns the parameter $s_1$ into a function of the quasi-momentum and the anisotropy.

The basic cell constructing one-magnon solutions is composed of a single $R$ matrix. In this case all the $G_k$ matrices are $2 \times 2$ upper triangular. The preservation of $U(1)$ symmetry further forces them to be diagonal. We choose the following ansatz 
\begin{equation}
    G_k= 
    \begin{pmatrix}
    1 & 0 \\
    0 & s_1 c_k
    \end{pmatrix} \ .
    \label{G1}
\end{equation}
The initial matrix $G_0$ defined in~\eqref{G0} is of this form with $c_0=1$. For $k>0$, the gauge freedom of the QR decomposition allows us to set the $c_k$ parameters to be real and positive. Equation~\eqref{rec2} translates then into the recursion relation
\begin{equation}
    c_k^2 =c_{k-1}^2 |s_2 |^2 + 1 \ ,
\end{equation}
for $k\geq 1$. Using the initial condition $c_0=1$, we arrive at the simple solution
\begin{equation}
c_k= \sqrt{k+1}\ ,
\end{equation}
where we have used that the quasi-momentum of the one-magnon solutions must be real, implying that $s_2$ is a phase.

Substituting into~\eqref{PfromG} and completing the matrix, we obtain the two-qubit quantum gates
\begin{equation}
P_k=
    \begin{pmatrix}
    1 & 0 &\hspace{-2mm}0&0\\[1mm]
    0 & {1 \over \sqrt{k+1}} & \hspace{-2mm}-{\sqrt{k  \over k+1}}\, e^{-\!ip} &0\\[2mm]
    0 & {\sqrt{k  \over k+1}}\, e^{ip} &\hspace{-2mm}{1 \over \sqrt{k+1}} &0\\[2mm]
    0 & 0 &\hspace{-2mm}0&1
    \end{pmatrix} \ ,
    \label{PM1}
\end{equation}
written with the convention
\begin{equation}
    \vcenter{\hbox{\includegraphics[width=0.85\columnwidth]{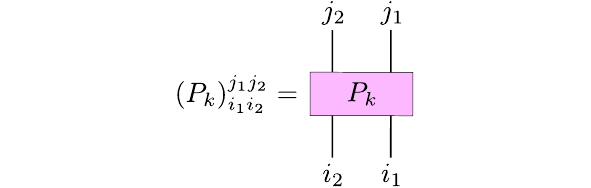}}} \ .
\label{eqn:P_2}
\end{equation}
We have ordered the input and output qubits from right to left instead of the customary left to right order, since this is better suited to the structure of the ABC. Indeed, with this convention the first two columns of~\eqref{PM1} describe $P_k\ket{0}$.

One-magnon solutions are just plane waves. The anisotropy parametrizes the strength of interactions among magnons and it hence should have no influence on these configurations.
Consequently, the unitaries $P_k$ are independent of $\Delta$.
This is in contrast with the ABA circuit network~\eqref{RTcircuit}, where every element depends on both $s_1$ and $s_2$, or equivalently, on $\Delta$ and $p$. It is only the output of the complete circuit that builds the one-magnon states which is independent of the anisotropy. In our scheme, the local dependence on $\Delta$ is confined to the unphysical matrices $G_k$.  
Our construction offers an economic way of implementing plane waves on quantum hardware, with just nearest-neighbor connectivity among qubits. The special case $p=0$ reproduces the optimal algorithm for obtaining the W state~\cite{2019_GH_W}.

\subsection{The XX model}
The XX model, obtained when $\Delta=0$, is the simplest member of the XXZ family. It describes a spinless free fermion system via the Jordan-Wigner transformation. This was used in~\cite{Verstraete_2009} to construct an efficient quantum circuit to prepare its eigenstates with a number of gates that is quadratic in the number of sites $N$. 
We will show that the ABC requires $\mathcal{O}(N M)$ gates.

We search for a decomposition of the unitaries $P_k$ in terms of two-qubit quantum gates. In particular, we will use the phased fSim ($F$) gate~\cite{2018Babush} as the basic building block of our circuit, defined as
\begin{equation}
   F = \left(\begin{array}{cccc}
    1 & 0 & 0 & 0 \\
    0 &  \cos \theta \, e^{i\alpha}& -\sin\theta \, e^{\!-i\beta} & 0 \\
    0 & \sin\theta \, e^{i\beta} & 
    \;\;\;  \cos\theta \, e^{\!-i\alpha} & 0 \\
    0 & 0 & 0 & 1 \\
\end{array}\right) \ .
\label{eqn:F}
\end{equation}
This matrix is a $U(1)$-symmetry-preserving generalization of the one-magnon unitaries~\eqref{PM1}. It can be implemented with five $R_X$ gates, five $R_Z$ gates and two CNOTs. We note that this unitary is a so called matchgate~\cite{jozsa2008matchgates}. Circuits consisting of matchgates are classically simulable and correspond to systems of free fermions. We use an ansatz in which the $F$ gates reproduce the same contraction structure as the $R$ matrices in the basic $\mathscr{R}_T$ cell, \ie
\begin{equation}
    \vcenter{\hbox{\includegraphics[width=0.85\columnwidth]{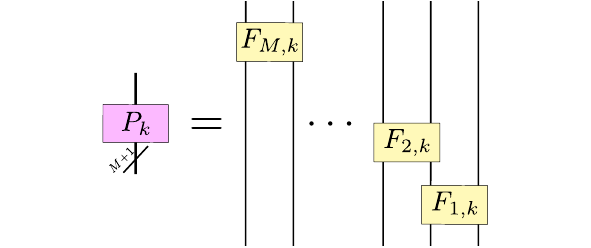}}} \ .
\label{eqn:ansatz}
\end{equation}
In Appendix~\ref{app:XX_model} we present the complete solution for two and three magnons in the XX model, including the closed-form of the matrices $P_k$. Furthermore, we verify the previous ansatz and analytically derive the parameters of the $F$ gates. 
We have checked numerically that~\eqref{eqn:ansatz} holds up to six magnons. It also holds for the unitaries $P_k$ with $k<M$, which act on a reduced number of qubits. Based on this evidence, we conjecture that it is valid in general for the XX model.

In spite of the similar decompositions of the matrices $P_k$ and ${\mathscr R}_T$, there is an important difference between them. Each $R_j$ matrix in ${\mathscr R}_T$
depends on a single magnon quasi-momentum $p_j$. On the contrary, the two-qubit gates $F_{j,k}$ contain information from all magnons down to position $j$, as seen in~\eqref{parI}-\eqref{parF}. Explicitly, the gate $F_{M,k}$ is a function of a single quasi-momentum $p_M$, while $F_{1,k}$ involves all of them.

The phase freedom shown in~\eqref{diagonals} is essential for~\eqref{eqn:ansatz} to hold.
It is simple to see that the one-spin-down sector of $P_k$ fixes completely this ansatz. The free phases affecting other symmetry sectors have then to be chosen appropriately for the so derived $F_{j,k}$ to describe the complete $P_k$. This choice has
been implemented in the two and three magnon unitaries presented in Appendix~\ref{app:XX_model}. Since $G_k$ and $P_k$ preserve the $U(1)$ symmetry, the one-spin-down sector can be determined independently from all others from the recursion relation~\eqref{rec2}. This implies that only ${M(M+1) \over 2}$ classical equations need to be solved. Hence our algorithm for the XX model is efficient both in its classical and quantum parts.

The total number of two-qubit gates for an ABC creating $M$ magnons on $N$ sites is $NM - {M(M+1) \over 2}$. All gates act upon nearest-neighboring qubits, which means that it is possible to implement it on quantum hardware with simple 1D connectivity. The total depth of the circuit is $N+M-1$, which represents a factor $\mathcal{O}(\log(N))$ improvement over the method presented in Ref.~\cite{Verstraete_2009}. Furthermore, it would be interesting to explore whether the depth of the ABC can be further reduced by considering additional connectivity~\cite{2019_GH_W}. 

It is worth noting that the circuit that we have obtained for the XX model is closely related to those considered in Refs.~\cite{2018Boixo, 2018Babush, arute2020observation} for free fermion models, that are based on preparing Slater determinants. These circuits have the same behavior in the depth and number of gates as the ABCs. This is to be expected since this model is non interacting and the eigenstates are themselves Slater determinants. There are sure to be interesting ties between the two approaches that may prove a fruitful research direction.

\subsection{The XXZ model}
The XXZ chain with non-vanishing anisotropy is an interacting model and thus computationally more challenging. Therefore, we expect the unitaries $P_k$ making up our ABC in the XXZ model to require a number of gates that grows exponentially with the number of magnons. Moreover, the number of equations that need to be solved in order to find the full matrices also scales exponentially with $M$. However, for sufficiently small numbers of magnons we find that this is not prohibitive.
Indeed, we can numerically obtain the matrices $P_k$ up to $M=12$ with modest computational resources. It should be noted that whether it is strictly necessary or not to calculate the entire matrices in order to find a quantum circuit that implements them is a matter of investigation. Likewise, whether an efficient quantum circuit for increasing $M$ may exist cannot be completely discarded based on our results.

For the decomposition of the unitaries $P_k$ we define the $\bar{F}$ gates identically to the $F$ gates~\eqref{eqn:F} but include an additional phase $e^{i\gamma}$ in the last diagonal element of the matrix. We note that the addition of this phase leads to the gates no longer being matchgates. Therefore, including this phase means the circuit can no longer be mapped to free fermions. This gate can be implemented with ten one-qubit gates and three CNOTs~\cite{PhysRevA.69.010301}. When $M=2$, we find that it is possible to decompose $P_k$ with
\begin{equation}
\vcenter{\hbox{\includegraphics[width=0.85\columnwidth]{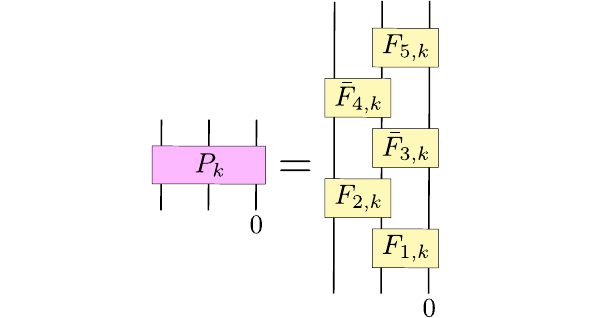}}} \ .
\end{equation}
The error in the compilation is measured using
\begin{equation}
    \epsilon=1-\left|\operatorname{Tr}\left(\la 0|U^{\dagger} P_{k}|0\ra\right)\right|^{2}/4^M \ ,
    \label{eq:compilation_error}
\end{equation}
achieving an order $10^{-10}$ precision. This renders the preparation of two-magnon configurations for small systems accessible to current quantum hardware.

For a larger number of magnons there is more freedom in how to distribute the $\bar{F}$ gates. 
As a first exploratory ansatz, we use the repetition of the structure in~\eqref{eqn:ansatz} in a layer-wise fashion, allowing for $\gamma$ phases in all gates. 

\begin{figure}[t!]
    \centering
    \includegraphics[width=\columnwidth]{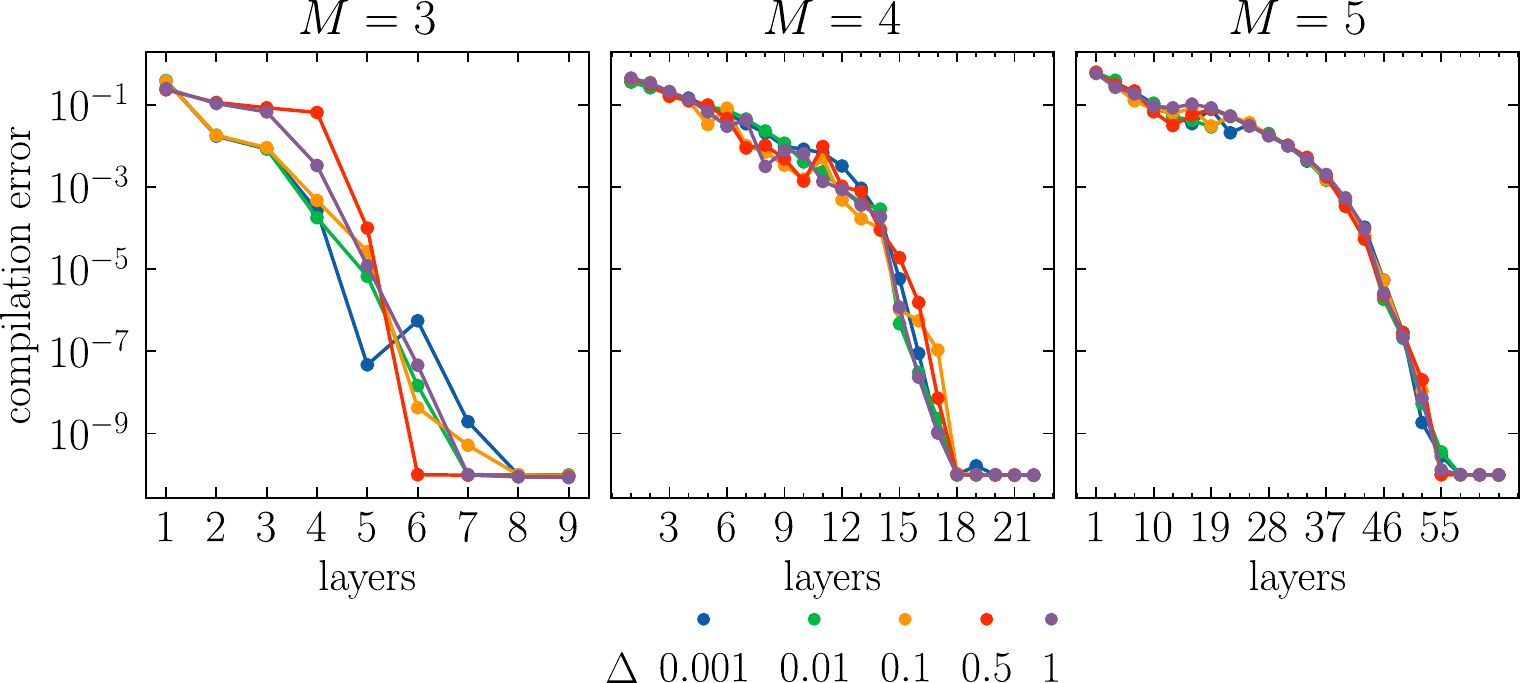}
    \caption{Compilation error~\eqref{eq:compilation_error} versus number of layers of the variational ansatz, for the isometry $P_{M}|0\ra$ and $M=3,4,5$ magnons. Each layer corresponds to the structure shown in~\eqref{eqn:ansatz}, but with $\bar{F}$ gates instead. We solved the Bethe equations for excited states with complex roots. Each point corresponds to the best result out of 5 optimizations with random initial parameters.}
    \label{fig:decomposition}
\end{figure}

We have verified up to 5 magnons that the compilation error associated with this ansatz does not depend on the matrix $P_k$ for $k\geq M$. Therefore, we have focused on $P_M$, {\it i.e.} on the unitary obtained right after the elimination of the ABA ancillary qubits.
Its numerical decomposition is addressed for several values of the anisotropy and $M=3,4,5$, as shown in Fig.~\ref{fig:decomposition}.
Our results indicate that a very small compilation error is obtained with around $8,18,55$ layers respectively, making up a total of $24,72,275$ $\bar{F}$ gates. The number of layers therefore appears to grow exponentially with the number of magnons for this ansatz. The compilation was achieved using the Berkeley Quantum Synthesis Toolkit (\texttt{BQSKit})~\cite{ed_younis_2022_6499836}, with a slight modification introduced to account for the fact that we are actually compiling $P_M|0\ra$ instead of the full unitary matrix. It should be stressed that these numerical results constitute a very preliminary study of the decomposition of the unitaries in the ABCs.

As a complementary direction, it is important to analyze the general properties of the unitaries $P_k$. With this aim we have derived the complete analytical solution for $M=2$ in the XXZ model (see Appendix~\ref{app:XXZ2_model}). We have seen that the first columns of $P_k$ reproduce the one-magnon solution~\eqref{PM1}. Also, we have checked that the same property holds for three magnons, and conjecture that it holds in general.
With this assumption, 
the solutions with $M' <M$ magnons turn out to be contained in those for $M$ magnons.
This property is evident for $\mathscr{R}_T$ and the $P_k$'s  from the XX model due to their single-layer structure of two-qubit matrices preserving the number of excitations. Since this simple structure is lost in the XXZ unitaries, we find remarkable that the hierarchy of solutions is preserved.

The simple structure of $\mathscr{R}_T$ and the XX model $P_k$'s has another important consequence. It forces many entries  of these matrices to vanish in spite of being compatible with the $U(1)$ symmetry. Starting from $M=3$, some of these entries become non-zero when the anisotropy does not vanish.
This is at the core of the difficulty encountered when decomposing the unitaries of the XXZ model.
A better understanding on the emergence of these entries should be crucial in searching for an optimal decomposition. Furthermore, the dependence on the index $k$ is an essential aspect of the ABCs, which deserves study both because of its theoretical and practical implications. A detailed investigation along these lines is left for future work.

Finally, it is interesting to note that the distinction between the XX model having a polynomial-depth circuit in both the number of qubits and the number of magnons, and the XXZ model likely having a circuit depth that scales exponentially with the number of magnons of the eigenstate, is matched by the dimension of the dynamical Lie algebras of the corresponding Hamiltonian generators. For the XX model, the dimension of this algebra is polynomial in the number of sites irrespective of the excitation subspace~\cite{cartan}. In contrast, for the XXZ model the dimension of the subspace algebra is exponential when the number of magnons scales linearly with the number of sites~\cite{larocca2021diagnosing}. Whether there exists a direct causal relation between these observations is left for now as an open question to be explored in future work.

\section{Unitary form of the Yang-Baxter equation}
\label{sec:unitary_YB}

The $R$ matrices of an integrable quantum system satisfy the celebrated YB equation
\begin{gather}
   \left(\mathbb{1}\otimes R(\lambda-\mu)\right)\left( R(\lambda)\otimes\mathbb{1}\right)\left(\mathbb{1}\otimes R(\mu)\right) = \label{eq:YB} \\[2mm]
   =  \left(R(\mu)\otimes\mathbb{1}\right)\left(\mathbb{1}\otimes R(\lambda)\right)\left(R(\lambda-\mu)\otimes\mathbb{1}\right)\nonumber\ ,
\end{gather}
where $\lambda$ and $\mu$ are the rapidity parameters introduced in Section~\ref{sec:ABA}. This equation is the statement that any process involving $N$ sites can be factorized into two-body pieces.
It guarantees that the BA state given in~\eqref{eqn:BA_state} is independent of the order of the parameters $\lambda_1,\dots, \lambda_M$. Let $\sigma$ be an arbitrary permutation of the $M$ rapidities. There exists then a $2^M \times 2^M$ tensor ${\mathscr R}_\sigma$ satisfying
\begin{equation}
        \vcenter{\hbox{\includegraphics[width=0.85\columnwidth]{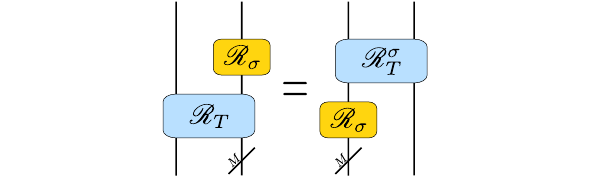}}} \ ,
    \label{eqn:Rsigma}
\end{equation}
where ${\mathscr R}_T^\sigma$ is defined by~\eqref{calR} with the permuted rapidities.
When substituting it  
into the network~\eqref{RTcircuit}, the independence on the ordering of the rapidities becomes manifest.
The simplest case of this relation is provided by the YB equation itself, whose graphic representation is 
\begin{equation}
        \vcenter{\hbox{\includegraphics[width=0.85\columnwidth]{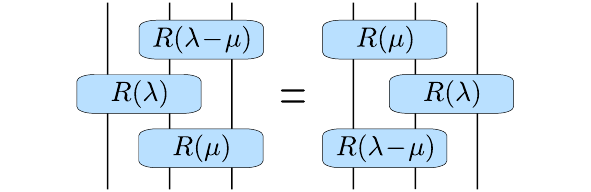}}} \ .
        \label{YBgraph}
\end{equation}
Here $\sigma$ permutes $\lambda$ and $\mu$, and ${\mathscr R}_\sigma=R(\lambda-\mu)$. For solutions containing more than two magnons, ${\mathscr R}_\sigma$ consists of a product of $R$ matrices depending on differences of rapidities. Although in general this product is not unique, the basic equation~\eqref{YBgraph} ensures that all choices lead to the same matrix ${\mathscr R}_\sigma$ (see Appendix~\ref{app:R_sigma}). 

We now analyze the implications of the YB equation on the quantum gates defining the circuit version of the ABA. Relation~\eqref{eqn:Rsigma} translates straightforwardly into
\begin{equation}
        \vcenter{\hbox{\includegraphics[width=0.85\columnwidth]{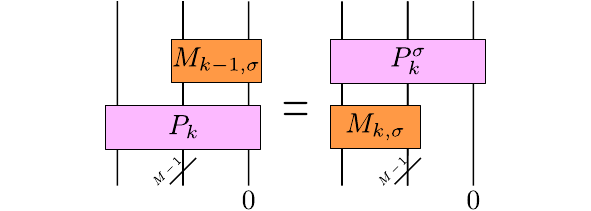}}} \ .
    \label{eqn:YB}
\end{equation}
The exchange matrices $M_k$ are obtained by dressing ${\mathscr R}_\sigma$ with the non-unitary pieces left by the QR decomposition that brings the ABA into unitary form
\begin{equation}
    M_{k,\sigma}=
    G_k^\sigma\,{\mathscr R}_\sigma\, G_k^{-1} \ ,
    \label{eqn:M}
\end{equation}
with $G_k^\sigma$ depending on the permuted rapidities. Moreover,~\eqref{eqn:Rsigma} also implies that the product $G_k^\sigma \, {\mathscr R}_\sigma$ satisfies the same recursion relations~\eqref{rec2} that define $G_k$. 
Both sets of solutions are related by 
\begin{equation}
    G_k^\dag \, G_k = {\mathscr R}_\sigma^\dag \, G_k^{\sigma  \dag}\, G^\sigma_k \, {\mathscr R}_\sigma  \ ,
    \label{uni}
\end{equation}
which is proven by induction, using that $G_0^\sigma \, {\mathscr R}_\sigma =G_0$.

It immediately follows that the exchange matrices $M_{k,\sigma}$ are unitary. Hence~\eqref{eqn:YB} is the unitary version of the YB equation.
It describes how the basic cell of the Bethe circuit changes under a permutation of the rapidity parameters. This reformulation of the YB equation allows for its test on quantum hardware with no restriction on the rapidities (see below).
It represents another application of the YB equation, and adds to those found in exactly-solvable models in Statistical Mechanics~\cite{baxter_exactly_1985} and the factorized $S$-matrices in relativistic quantum field theory models~\cite{ZZ79}.

The exchange matrices for $M=2$ are calculated from the direct relation with the $R$ matrix 
\begin{equation}
       \vcenter{\hbox{\includegraphics[width=0.85\columnwidth]{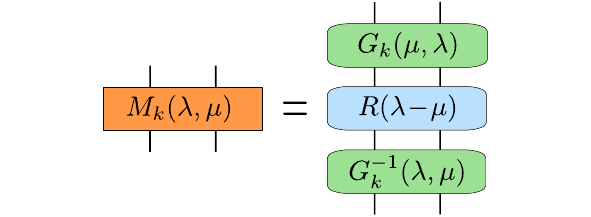}}} \ .
       \label{MM2}
\end{equation}
We have dropped the sub-index $\sigma$ in $M_k$ since in this case there is only one possible permutation, which interchanges $\lambda$ and $\mu$. The analytical expression is presented in Appendix~\ref{exchange}. 
Although all elements on the right-hand side of~\eqref{MM2} depend on the anisotropy,
their product does not. This implies that $M_k$ is the same function of the magnon quasi-momenta for all members of the XXZ family. This shows that the matrices~\eqref{MM2} cannot be interpreted as unitary versions of the $R$ matrix. 
We stress that the exchange matrices depend on the anisotropy for $M\geq3$. These matrices are interesting objects in their own right, deserving further study.
 
\begin{figure*}[t!]
    \centering
    \includegraphics[width=2\columnwidth]{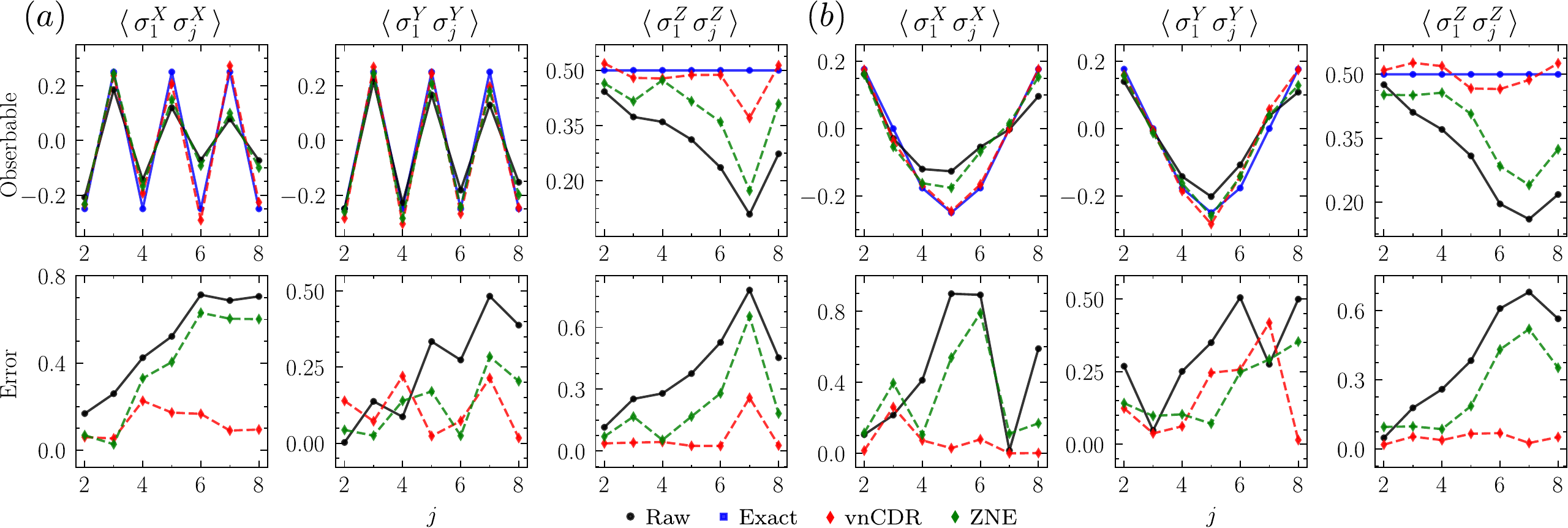}
    \caption{Two-point correlators calculated using the quantum computer \textit{IBM\_Montreal}. Qubits are numbered from right to left, i.e. $j=1$ indicates the rightmost qubit. We simulate one plane wave with $N=8$ sites. In $(a)$ $p=\pi$, in $(b)$ $p = 5.498$. The first row shows the correlators, while the second row shows their weighted error, defined in Eq.~\ref{eqn:error}. Error mitigation is implemented with ZNE and vnCDR. The raw observables were calculated to have a mean weighted error of $[(a)\text{ }0.38, (b)\text{ }0.38]$, ZNE reduced this error to $[(a)\text{ }0.24, (b)\text{ }0.25]$ and vnCDR to $[(a)\text{ }0.10, (b)\text{ }0.09]$.
    }
    \label{fig:correlators_one_magnon}
\end{figure*}

 \begin{figure*}[ht!]
    \centering
    \includegraphics[width=2\columnwidth]{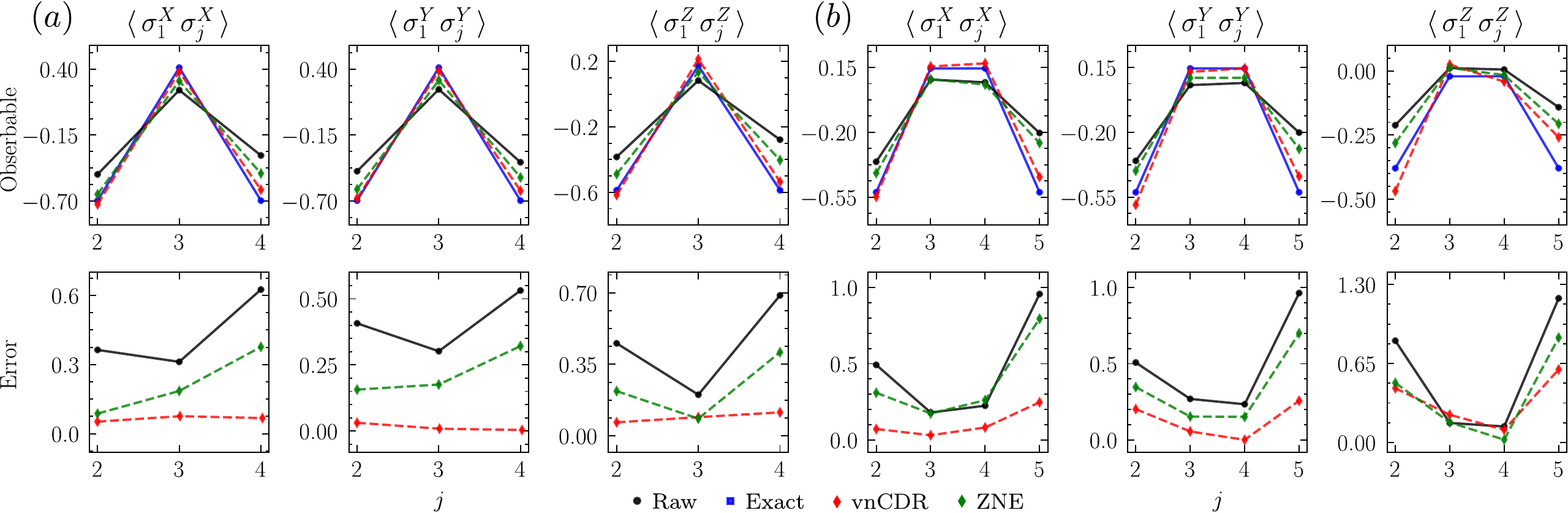}
    \caption{Two-point correlators for several states with $2$ magnons calculated using the quantum computers  \textit{IBM\_Montreal} in (a) and \textit{IBM\_Mumbai} in (b).  In $(a)$ we prepare the ground state of the $XXZ$ model for $4$ sites and $\Delta=0.5$. In $(b)$ we prepare an excited state of the $XX$ model with $2$ magnons and $5$ sites. The first row shows the correlators, while the second row shows their weighted error, defined in Eq.~\ref{eqn:error}. The raw observables were calculated to have a mean weighted error of $[(a)\text{ }0.43, (b)\text{ }0.51]$, ZNE reduced this error to $[(a)\text{ }0.22, (b)\text{ }0.37]$ and vnCDR to $[(a)\text{ }0.06, (b)\text{ }0.19]$.
    }
    \label{fig:correlators2mag}
\end{figure*}
\section{Numerical and Experimental Results}
\label{sec:experiments}

We performed numerical simulations to verify our theoretical results, using the open-source library \texttt{Qibo}~\cite{efthymiou2021, efthymiou2021qiboteam/qibo}. We numerically solved the Bethe equations~\eqref{eqn:Bethe_eqs} and simulated the unitary circuits at the level of the $P_k$ matrices for several excited states of the XXZ model up to 24 sites and 12 magnons. We compared the resulting eigenstates with the simulated ABA states, finding a perfect agreement between the two. The circuits were simulated in double precision using the \texttt{qibojit} backend~\cite{efthymiou2022qiboteam/qibojit} on multi-threading CPU. They were directly obtained by computing the QR decompositions that are required to convert the ABA into a deterministic quantum circuit. The programs to reproduce such simulations can be found at~\cite{ABC}.

\subsection{Plane waves on quantum hardware}

We also implemented the circuit construction for plane wave states for a system size $N=8$ on the quantum computer \textit{IBM\_Montreal}. We explored the two-point correlators from the prepared state (see Fig.~\ref{fig:correlators_one_magnon}).

As previously shown, the 2 qubit $P_{k}$ unitary gates are simply phased fSim gates when $M=1$. These gates were decomposed into the IBM native gate set to be implemented on the hardware. The circuits used to prepare these states have total depth 65. Each unitary gate involved $2$ CNOT gates, leading to $2(N-1)$ CNOT gates in total.

Current devices suffer from significant hardware noise. In order to obtain the best possible results it is necessary to use error mitigation which focuses on reducing the impact of noise. There are many current error mitigation approaches and unifications thereof, each with their respective advantages and disadvantages~\cite{bultrini2021unifying, cirstoiu2022volumetric}. Exploring how error mitigation performs for a task of interest on real quantum computers give insight onto their performance in realistic scenarios.

In this work we implemented three techniques: zero-noise extrapolation (ZNE)~\cite{zne}, Clifford data regression (CDR)~\cite{cdr} and variable noise Clifford data regression (vnCDR)~\cite{vncdr}. We used the open source software package \texttt{Mitiq}~\cite{mitiq} to execute these methods. For more details regarding the implementation of these techniques we refer the reader to Appendix~\ref{app:mitigation}. 

We benchmark the performance of each method by calculating a weighted error, defined as:
\begin{equation}
    \langle O_j \rangle_{Err} = \frac{\abs{\langle O_j\rangle_{exp} - \langle O_j\rangle_{exact}}}{\operatorname{mean}\left(\abs{\langle O_{j} \rangle_{exact}}\right)}
\label{eqn:error}
\end{equation}
where $O_{j}$ is some observable of interest and $\langle O_{j} \rangle_{exp}$ is the estimated value for that observable obtained experimentally with or without error mitigation. The mean is taken over the different qubit positions. This definition prevents the error from diverging if $\langle O_j\rangle_{exact}$ approaches zero while allowing the errors corresponding to several observables to be compared. To simplify the comparison between mitigation techniques we average the above error metric across all the different observables. We expect this metric to reflect the overall performance of the mitigation methods while also taking into account the magnitude of the observables to mitigate. Furthermore, to simplify the presentation we omit the CDR mitigated results from our plots, we see that in general the performance is significantly better than ZNE but worse than vnCDR. 

Error mitigation significantly improves the results obtained from the real device (see Fig.~\ref{fig:correlators_one_magnon}). As the observables become less local and the circuit depth increases, the quality of the raw and error mitigated results tends to decrease. This is particularly noticeable in the results obtained for the $\langle \sigma^{Z}_{1}\sigma^{Z}_{j}\rangle$ correlators. Clearly both vnCDR and ZNE tend to perform best in shallower, less noisy circuits although they still improve results for the deepest circuits explored here.

\subsection{Two-magnon states on quantum hardware}
In addition, we implemented the circuits for two magnon states on current quantum hardware. We constructed the ground state of the $XXZ$ model for $4$ sites and an excited state of the $XX$ model for $5$ sites. We constructed the ground state of the $XX$ model for $4$ sites in Appendix~\ref{app:XX_GS}.

The circuit for the ground state of the $XXZ$ model consists of $5$ $F$ gates and $2$ $\bar{F}$ gates, structured as follows:
\begin{equation}
        \vcenter{\hbox{\includegraphics[width=0.85\columnwidth]{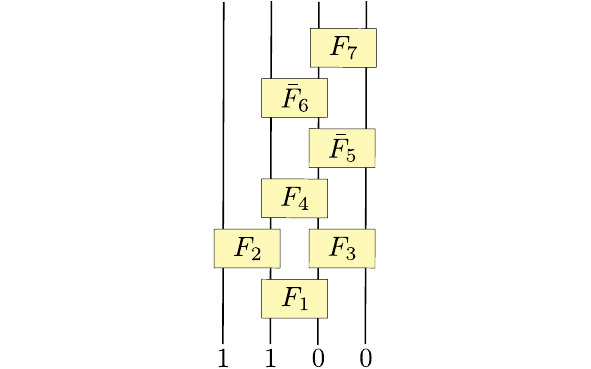}}} \ .
\end{equation}
We only need to determine the action of the last unitary $P_{N-1}$ on the state $|011\rangle$ and this can be achieved with a single layer of $F$ gates. 

Once the $F$ and $\bar{F}$  gates have been decomposed into the IBM native gate set the circuit to prepare these states have depths $57$ and $49$ and involve $16$ and $14$ CNOT gates respectively. 

We evaluated the two-point correlators for these states across the chain, and mitigated the effect of hardware noise with ZNE and vnCDR (see Fig.~\ref{fig:correlators2mag}). The mitigation methods improve upon the raw data. For two magnon states we explored smaller system sizes due to the increased scaling of depth with the number of qubits. We find that good agreement can be obtained between the exact and error mitigated observables, with vnCDR reducing the effect of noise the most.

Overall, these experiments show a proof of principle implementation of our approach for low numbers of magnons. Furthermore, they highlight the utility of error mitigation. In particular they show further evidence that learning based error mitigation is practically useful in reducing the effects of hardware noise. For larger scale implementations a combination of noise reduction and error mitigation techniques will be needed, which presents an exciting challenge for future experiments.

\subsection{Yang-Baxter equation on quantum hardware}
The significance of the YB equation has motivated its experimental implementation. A two-dimensional version of the YB equation has been verified optically~\cite{Zheng:13} and by nuclear magnetic resonance~\cite{vind_experimental_2016}. We have also verified the unitarised YB equation~\eqref{eqn:YB} on the cloud quantum computer \textit{IBM\_Cairo}, for the case $M=2$ magnons in the XX model, given by
\begin{equation}
        \vcenter{\hbox{\includegraphics[width=0.85\columnwidth]{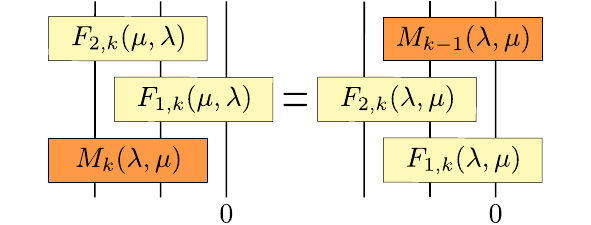}}} \ .
\label{eqn:YB_experimental}
\end{equation}
We used state tomography to compute the density matrices associated with the right ($\rho_r$) and left ($\rho_l$) output states at both sides of~\eqref{eqn:YB_experimental} (see Table~\ref{tab:YB}). We computed these density matrices for each of the four possible initial input states, and we did so for the $P_2$ gate with Bethe roots $\lambda_1=-1/\sqrt{3}$ and $\lambda_2=1/\sqrt{3}$. In order to compare $\rho_r$ and $\rho_l$ we determined the fidelity, given by
\begin{equation}
    F=\left(\operatorname{Tr}\sqrt{\sqrt{\rho_r}\rho_l\sqrt{\rho_r}}\right)^2 \ .
\end{equation}
This heralds the first implementation of the YB equation on a superconducting quantum computer. 
\begin{table}
\centering
\renewcommand{\arraystretch}{1.3}
\setlength{\tabcolsep}{7pt}
\begin{tabular}{cc}
Initial sate & Fidelity\\
\hhline{==}
$|000\rangle$ & 0.969\\
$|010\rangle$ & 0.964\\
$|100\rangle$ & 0.962\\
$|110\rangle$ & 0.950\\
\end{tabular}
\caption{\label{tab:YB}Verification of the unitarised YB equation using the   \textit{IBM\_Cairo} cloud quantum computer. The left column shows the initial state fed into~\eqref{eqn:YB}, and the right column shows the fidelity between the output states at both sides of this equation.}
\end{table}

\section{Discussion}
\label{sec:discussion}
In this work, we have introduced a method to exactly prepare eigenstates of quantum integrable vertex models on programmable digital quantum computers. Our approach relies on using the QR decomposition as the main tool to bring the ABA to unitary form. In contrast to previous proposals, our method works for both real and complex roots of the Bethe equations and is deterministic. Both the circuit depth and gate complexity of our approach scale linearly with the number of qubits. However, we expect an exponential scaling in the number of magnons in general. This could affect the classical preprocessing, the required compilation step, and the circuit depth.

Despite this, we find that with modest classical computational resources one can obtain a unitary circuit representation for interesting states. For the XX model, which can be mapped to free fermions, we find an efficient gate decomposition with polynomial classical effort. In particular, our approach produces quantum circuits that match the state-of-the-art $\mathcal{O}(N)$ depth of~\cite{2018Boixo, 2018Babush, arute2020observation, kokcu2021algebraic}.

Our algorithm opens up the possibility to prepare highly non-trivial ABA eigenstates on quantum computers. Foreseeable applications include using these states to study Hamiltonian quenches that may be inaccessible to classical methods. More generally, these states could be used as inputs to other quantum algorithms. For instance, it would be interesting to explore if they can be used to initialize variational quantum algorithms. Using such states may provide an initial state with sufficient overlap with the desired output for the optimization to be successful. This would combat the trainability issues of such approaches~\cite{2019_initialization_BP,CF_2021}.
Furthermore, our algorithm could be used to benchmark quantum hardware on strongly-correlated states whenever analytical solutions are known for some expectation values. This can be thought of as a type of application-oriented benchmark~\cite{lubinski2021applicationoriented}.

There remain many open questions that would be interesting to explore in future works. A clear next step will be to investigate the optimal strategy with which to compile the unitary gates $P_{k}$ to improve the performance for systems of many magnons~\cite{kiani2020learning,larocca2021theory}. 
Extending our method to open boundary conditions and to other models, such as the Hubbard or Kondo models, is another clear research direction. Additionally, we want to stress that our approach can be used to represent an MPS as a deterministic quantum circuit, which would enable direct preparation of these states on a quantum computer.

\section*{Acknowledgements}
We acknowledge Luigi Amico, Marco Cerezo, Lukasz Cincio, Artur Garcia-Saez, Karen Hallberg, Hosho Katsura, Martín Larocca, José Ignacio Latorre, Rafael Nepomechie and Balázs Pozsgay for useful discussions. We thank the IBM Quantum team for making devices available via the IBM Quantum Experience. The access to the IBM Quantum Experience has been provided by the CSIC IBM Q Hub. A.S. is supported by the Spanish Ministry of Science and Innovation under grant number SEV-2016-0597-19-4. M.H.G. was supported by “la Caixa” Foundation (ID 100010434), Grant No. LCF/BQ/DI19/11730056 and by the U.S. DOE, Office of Science, Office of Advanced Scientific Computing Research, under the Quantum Computing Application Teams program. D.G.M. is supported by project QuantumCAT (ref.  001- P-001644), co-funded by the Generalitat de Catalunya and the European Union Regional Development Fund within the ERDF Operational Program of Catalunya. This work has also been financed by the Grant PGC2018-095862-B-C21 funded by MCIN/AEI/10.13039/501100011033 and by  “ERDF A way of making Europe”,
by the Spanish Research Agency (Agencia
Estatal de Investigación) through the Grant IFT Centro de Excelencia Severo
Ochoa No CEX2020-001007-S, funded by MCIN/AEI/10.13039/501100011033, by
the Madrid grant PGC2018-095862-B-C21, QUITEMAD+ S2013/ICE-2801, and by the CSIC Research Platform on Quantum Technologies PTI-001.
\bibliographystyle{unsrtnat}
\bibliography{main.bib}

\clearpage
\appendix
\counterwithin{figure}{section}
\section{Non-unitarity of the \texorpdfstring{$R$}{R} matrices}
\label{app:R_non_unitary}

We show here that the $R$ matrices~\eqref{eqn:R} appearing in the Algebraic Bethe Ansatz (ABA) are not unitary in general. We start with
\beq RR^\dagger = |\rho|^2 \left(\begin{array}{cccc}       1 & 0 & 0 & 0 \\
      0 & |s_1|^2 + |s_2|^2 & s_1 s_2^* +s_2 s_1^* & 0 \\
      0 & s_1 s_2^* +s_2 s_1^* & |s_1|^2 + |s_2|^2 & 0 \\
      0 & 0  & 0 & 1
\end{array} \right) \,,\eeq
where $^*$ denotes the complex conjugate. In order for $R$ to be unitary, the following conditions must be satisfied,
\beq \begin{split}  |\rho|^2 = 1\,, \\  |s_1|^2 + |s_2|^2 = 1 \,, \\ s_1 s_2^* +s_2 s_1^*  = 0 \,. \end{split} \eeq

After some straightforward algebra, the last two conditions are seen to imply
\begin{equation}
    \Im \lambda = 1 + {2 \pi n \over \gamma} ,
\end{equation}
with $n$ an integer. However, in general the parameters $\lambda$ that solve the Bethe equations do not satisfy this requirement.

\section{The monodromy and transfer matrices}
\label{app:integrable}
In the main text we have used a matrix $R_{12}:  {\cal H}_1 \otimes {\cal H}_2 \rightarrow  {\cal H}_2 \otimes {\cal H}_1$, where
${\cal H}_1$ and ${\cal H}_2$ are the vector space $\mathbb{C}^{\otimes 2}$. To define a transfer matrix it is customary to employ a matrix ${\cal R}_{12}:  {\cal H}_1 \otimes {\cal H}_2 \rightarrow  {\cal H}_1 \otimes {\cal H}_2$, that is related to
$R_{12}$ by a permutation  $P_{12}$, 
\beq
{\cal R}_{12}  = {P}_{12} \, {R}_{12} \, ,
\eeq
that reads in components
\begin{equation}
       \vcenter{\hbox{\includegraphics[width=0.85\columnwidth]{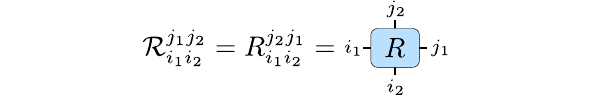}}} \ .
       \label{checkR}
\end{equation} 
The YB equation for the matrix $R(\lambda)$, given in~\eqref{eq:YB}, 
can be expressed in terms of  ${\cal R}(\lambda)$ as 
\beq
{\cal R}_{12}(\lambda- \mu)  {\cal R}_{13}(\lambda)   {\cal R}_{23}(\mu) = {\cal R}_{23}(\mu)  {\cal R}_{13}(\lambda)   {\cal R}_{12}(\lambda- \mu)  \ .
\label{YB2}
\eeq
The monodromy matrix $T(\lambda)$  is a linear map from ${\cal H}_a \otimes {\cal H}_1 \otimes \dots \otimes {\cal H}_N$ to itself,
where  ${\cal H}_a$ is an auxiliary space and ${\cal H}_j$ the local  quantum spaces $(j=1, \dots, N)$.
Its definition is~\cite{gomez_quantum_1996} 
\beq
{T}(\lambda) = {\cal R}_{a N}(\lambda) {\cal R}_{a N-1}(\lambda)   \dots {\cal R}_{a 2}(\lambda)  {\cal R}_{a 1}(\lambda)  \ .
\label{eq:TransferMatrix}
\eeq
where 
\beq
{\cal R}_{a j}(u) : {\cal H}_a \otimes {\cal H}_j \rightarrow  {\cal H}_a \otimes {\cal H}_j \ , \quad j=1, \dots, N  \, .
\eeq
Equation~\eqref{eq:TransferMatrix} reads in components
\beq
(T_a^b)_{i_1, \dots, i_N}^{j_1, \dots, j_N} =  \sum_{a_1, \dots, a_{N-1}} {\cal R}_{a_{N-1} i_N}^{b j_N} \dots {\cal R}_{a_1 i_2}^{a_2 j_2} \, 
 {\cal R}_{a i_1}^{a_1 j_1}   \ . 
\label{eq:TM_1}
\eeq  
Notice that the contraction of the matrices ${\cal R}_{a j}$  in the auxiliary space follows the rule  $(X Y)^a_c = X^a_b Y^b_c$, so that upper and lower  indices correspond to row and column indices respectively. For the XXZ model we are considering $T(\lambda)$ is the  $2 \times 2$ operator matrix
\beq
\label{mono}
T(\lambda) = \left(
\begin{array}{cc}
T_0^0 & T_1^0 \\
T_0^1 & T_1^1 \\
\end{array}
\right)  =  \left(
\begin{array}{cc}
A & B \\
C & D  \\
\end{array}
\right) \ . 
\eeq
Equation~\eqref{eq:TM_1} can be given the tensor network notation by simply permutating the upper indices of the matrices ${\cal R}$, \beq
(T_a^b)_{i_1, \dots, i_N}^{j_1, \dots, j_N} =  \sum_{a_1, \dots, a_{N-1}} {R}_{a_{N-1} i_N}^{j_N b} \dots {R}_{a_1 i_2}^{j_2 a_2 } \, 
 {R}_{a i_1}^{j_1 a_1}   \ . 
\eeq
whose graphical representation is
\begin{equation}
       \vcenter{\hbox{\includegraphics[width=0.85\columnwidth]{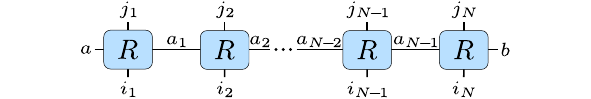}}} \ .
\end{equation} 
The case $a=1$ and $b=0$ reproduces~\eqref{B} giving the operator $B(\lambda)$. Finally, the transfer matrix is defined as the trace of the monodromy matrix~\eqref{mono} in the auxiliary space which yields 
\begin{equation}
    t(\lambda) = \sum_{a=0,1} T_a^a(\lambda) = A+ D \ .
    \label{transfer}
\end{equation}

\section{General solution for two magnons}
\label{app:XXZ2_model}

The transformation of the ABA into a quantum circuit can be carried out  analytically for the case of two magnons and for any number of sites $N$ . 

The ${\mathscr R}_T$ basic cell is 
\begin{equation}
       \vcenter{\hbox{\includegraphics[width=0.85\columnwidth]{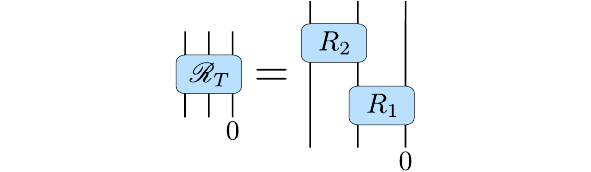}}} \ ,
       \label{RT2}
\end{equation}
with $R_1$ and $R_2$ parametrized by $r_{1,2}$ and $s_{1,2}$ respectively.
The matrix $G_0$~\eqref{G0}, which defines the starting point of our algorithm, is given by
\begin{equation}
G_0=\!\begin{pmatrix}
1&0&0&0\\[1mm]
0&s_1  & r_1 s_2 &0
\end{pmatrix} .
\label{G00}
\end{equation}
From now on,
we follow the right to left ordering of qubits explained in~\eqref{eqn:P_2}.
Notice that this is opposite to the customary left to right order used in Appendix~\ref{app:integrable}. Namely
\begin{equation}
\vcenter{\hbox{\includegraphics[width=0.85\columnwidth]{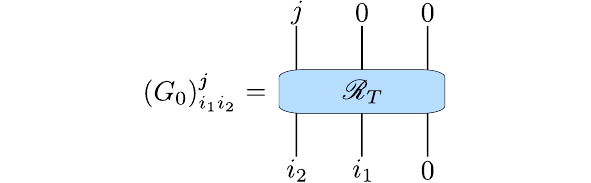}}} \ .
\label{eqnnn:G00}
\end{equation}
When writing down the entries of an ABC matrix, we will always adopt this convention.
Substituting into the recursion relations~\eqref{rec2}, we find
\begin{equation}
\hspace{-5mm}\; \;G_k=\!\begin{pmatrix}
1&0&0&\hspace{-8mm}0\\[1mm]
0&c_k s_1  &\!c_k r_1 \!\left(s_2\!+\!{f_k (s_1^2+(r_2\!-\!s_2)s_2)s_2^\ast  \over c_k^2 } \right)&\hspace{-8mm}0\\[2mm]
0& 0& {-d_k r_1(s_1^2+(r_2\!-\!s_2)s_2) \over c_k} &\hspace{-8mm}0\\[1mm]
0&0&0& \hspace{-15mm}- e_k  r_1 s_1(1\!+\!r_2 s_2)
\end{pmatrix} ,
\end{equation}
for $k\geq 1$. The coefficients
$c_k$, $d_k$ and $e_k$ are assumed to be real and positive using the gauge freedom~\eqref{diagonals}. The new coefficients satisfy 
\begin{eqnarray}
c_k^2 &=&c_{k\!-\!1}^2 |s_2 |^2 + 1 \label{ck} \ , \label{ccoeff}\\[1mm]
f_k &= & f_{k\!-\!1} r_2 s_2^\ast+c_{k\!-\!1}^2  
 \ , \\[1mm] c_{k\!-\!1}^2 d_k^2& =& c_k^2 \, d_{k\!-\!1}^2 |r_2|^2 +|f_k|^2  \ ,
\label{dcoeff}
\end{eqnarray}
while $e_k$ is given by
\begin{eqnarray}
e_k^2 &=&d_k^2 -{2 \Delta \over |1\!+\!r_2s_2|^2} \Big[ (d_k^2\!-{\bar e}_k^2)(1\!+\!r_2 s_2) s_2^\ast  \\ & +& (d_k^2\!-{\bar e}_k^{\ast 2})(1\!+\! r_2 s_2)^{\! \ast} s_2 \Big]  + \,{4 \Delta^2   |s_2 r_2 |^2 \over |1\!+\!r_2s_2|^2} \; d_{k\!-\!1}^2  \ ,
\nonumber
\label{ehorror}
\end{eqnarray}
with
\begin{equation}
{\bar e}_k^2= {\bar e}_{k\!-\!1}^2 |r_2 s_2 |^2+  f_k   \ .
\label{ebarcoeff}
\end{equation}
The initial conditions for these equations are
\begin{equation}
    c_0=1 \ , \hspace{5mm} f_0=d_0=e_0=0 \ .
    \label{icond}
\end{equation}
We obtain the first, 2-qubit unitary of the circuit from equation~\eqref{PfromG2}
\begin{equation}
P_1=\begin{pmatrix}
1&0&0&0\\[1mm]
0&{1 \over c_1}  &{s_2^\ast \over c_1}&0\\[2mm]
0&{s_2 \over c_1}& {-1 \over c_1} &0\\[1mm]
0&0&0& 1
\label{firstP}
\end{pmatrix} .
\end{equation}
Equation~\eqref{PfromG} determines the  action of 3-qubit unitaries  $P_{k>1}$ on the state $\ket{0}$ to be 
\begin{equation}
\hspace{-5mm} P_k \!\ket{0}=\!
\begin{pmatrix}
1&0&0&0 \\
0&{1 \over c_k}  &{f_k  s_2^\ast \over c_k d_k} &0 \\[2mm]
0& {c_{k\!-\!1} s_2  \over c_k}&
{-f_k \over c_{k\!-\!1} c_k d_k}&0\\[1mm]
0&0&0&\hspace{-2mm}{-1 \over c_{k\!-\!1} e_k} \! \left(\!  f_k \!-\!{2 \Delta r_2 \over 1 + r_2 s_2} |s_2|^2 f_{k\!-\!1} \!  \right) \!\!\\[1mm]
0&0&{c_k d_{k\!-\!1}r_2 \over c_{k\!-\!1} d_k}&0\\[1mm]
0&0&0&{d_{k\!-\!1} r_2 \over  c_{k\!-\!1} e_k} \! \left(\!1 - {2 \Delta s_2 \over 1 + r_2 s_2}\! \right)\\[3mm]
0&0&0&{e_{k\!-\!1}r_2 s_2  \over e_k}\\[1mm]
0&0&0&0
\end{pmatrix}
\label{P2} \ .
\end{equation}
The recursion relation~\eqref{ck} defining the coefficients $c_k$ is easily solved
\begin{equation}
    c_k=\sqrt{1+|s_2|^2+ \dots+ |s_2|^{2k}} \ .
\end{equation}
\noindent Recalling that $s_2$ is just a phase for one-magnon solutions, we obtain that the first columns of $G_k$  and $P_k \ket{0}$ for two magnons contain the one-magnon solution~\eqref{G1} and the corresponding $P_k \ket{0}$ part of~\eqref{PM1}.

\section{The XX model}
\label{app:XX_model}

The XX model is obtained when 
$\Delta=0$. It is a free system
and thus implies drastic simplifications with respect to the general XXZ model. The two magnon solution~\eqref{P2} reduces in the XX model to
\begin{equation}
P_k\!\ket{0}=
\begin{pmatrix}
1&0&0&0 \\
0&{1 \over c_k}  &{f_k  s_2^\ast \over c_k d_k} &0 \\[2mm]
0& {c_{k\!-\!1} s_2  \over c_k}&
{-f_k \over c_{k\!-\!1} c_k d_k}&0\\[1mm]
0&0&0&\hspace{-2mm}{-f_k \over c_{k\!-\!1} d_k}\\[1mm]
0&0&{r_2 c_k d_{k\!-\!1} \over c_{k\!-\!1} d_k}&0\\[1mm]
0&0&0&{d_{k\!-\!1} r_2 \over  c_{k\!-\!1} d_k} \\[3mm]
0&0&0&{r_2 s_2 d_{k\!-\!1} \over d_k}\\[1mm]
0&0&0&0
\end{pmatrix}
\label{P20} \ .
\end{equation}

We have also derived the explicit three-magnon solution. The ${\mathscr R}_T$ basic cell is now
\begin{equation}
       \vcenter{\hbox{\includegraphics[width=0.85\columnwidth]{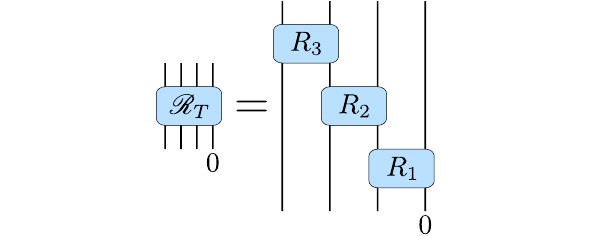}}}\ ,
       \label{MM3}
\end{equation}
with $R_1$, $R_2$ and $R_3$ parametrized by $t_{1,2}$, $r_{1,2}$ and $s_{1,2}$ respectively.
We will just sketch the first iterations.
The initial matrix $G_0$~\eqref{G0} has now dimension $2 \times 8$. The first step~\eqref{P1G_1} results in a $4 \times 8$ non-unitary $G_1$ and the $4 \times 4$ gate $P_1$~\eqref{firstP}. The second iteration leads to $G_2$ and $P_2$ both of dimension $8 \times 8$. From then on each iteration distill a four qubit unitary, whose action on the rightmost $\ket{0}$ is
\begin{widetext}
\begin{equation}
\hspace{-15mm}P_k \! \ket{0}=
\begin{pmatrix}
1&0&\hspace{-2mm}0&\hspace{-4mm}0&\hspace{-4mm}0&\hspace{-6mm}0&0&\hspace{-3mm}0 \\[2mm]
0& {1 \over c_k} & \hspace{-2mm}{f_k s_2^\ast \over c_k d_k} &\hspace{-4mm}0& \hspace{-4mm}{h_k s_2^\ast r_2^\ast \over d_k g_k} &\hspace{-6mm}0&0&\hspace{-3mm}0 \\[2mm]
0& \! {c_{k\!-\!1} s_2 \over c_k} &\hspace{-2mm}{-f_k \over c_{k\!-\!1} c_k d_k} &\hspace{-4mm}0&\hspace{-4mm} {-h_k r_2^\ast \over c_{k\!-\!1}d_k g_k}&\hspace{-6mm}0&0&\hspace{-3mm}0 \\[2mm]
0&0&\hspace{-2mm}0& \hspace{-4mm}{-f_k \over c_{k\!-\!1} d_k} &\hspace{-4mm}0& \hspace{-5mm}{-c_k h_k r_2^\ast \over c_{k\!-\!1}d_k g_k}&0&\hspace{-3mm}0\\[2mm]
0&0& \hspace{-2mm}{c_k d_{k\!-\!1} r_2 \over c_{k\!-\!1} d_k}&\hspace{-4mm}0&\hspace{-4mm} {-h_k f_k^\ast \over c_{k\!-\!1} d_{k\!-\!1}d_k g_k}&\hspace{-5mm}0&0&\hspace{-3mm}0\\[2mm]
0&0&\hspace{-2mm}0&\hspace{-3mm} {d_{k\!-\!1} r_2 \over c_{k\!-\!1} d_k}&\hspace{-4mm}0& \hspace{-7mm}{-h_k f_k^\ast \over c_{k\!-\!1} d_{k\!-\!1}c_k d_k g_k}& \hspace{-2.5mm}{-c_{k\!-\!1} h_k s_2^\ast \over d_{k\!-\!1}c_k g_k} &\hspace{-3mm}0\\[2mm]
0&0&\hspace{-2mm}0&\hspace{-4mm} {d_{k\!-\!1} s_2 r_2 \over d_k}&\hspace{-4mm}0& \hspace{-7mm}{-h_k f_k^\ast s_2 \over d_{k\!-\!1}c_k d_k g_k}& \hspace{-2.5mm}{ h_k  \over d_{k\!-\!1}c_k g_k} &\hspace{-3mm}0\\[2mm]
0&0&\hspace{-2mm}0&\hspace{-4mm}0&\hspace{-4mm}0&\hspace{-6mm}0&0&\hspace{-5mm}
{h_k \over d_{k\!-\!1} g_k} \! \\[2mm]
0&0&\hspace{-2mm}0&\hspace{-4mm}0&\hspace{-4mm}{d_k g_{k\!-\!1} t_2 \over d_{k\!-\!1} g_k} &\hspace{-5mm}0&0&\hspace{-3mm}0 \\[2mm]
0&0&\hspace{-2mm}0&\hspace{-4mm}0&\hspace{-4mm}0&\hspace{-5mm}{d_k g_{k\!-\!1} t_2 \over d_{k\!-\!1} c_k g_k} & \hspace{-1mm} {f_k g_{k\!-\!1} s_2^\ast t_2 \over d_{k\!-\!1} c_k g_k} & \hspace{-3mm} 0\\[2mm]
0&0&\hspace{-2mm}0&\hspace{-4mm}0&\hspace{-4mm}0&\hspace{-7mm}{c_{k\!-\!1} d_k g_{k\!-\!1} s_2 t_2 \over d_{k\!-\!1} c_k g_k} & \hspace{-1mm} {-f_k g_{k\!-\!1}  t_2 \over c_{k\!-\!1} d_{k\!-\!1} c_k g_k} & \hspace{-3mm} 0 \\[2mm]
0&0&\hspace{-2mm}0&\hspace{-4mm}0&\hspace{-4mm}0&\hspace{-5mm}0 &0&\hspace{-6mm}{-f_k g_{k\!-\!1}  t_2 \over c_{k\!-\!1} d_{k\!-\!1} g_k}\!\!\!\\[2mm]
0&0&\hspace{-2mm}0&\hspace{-4mm}0&\hspace{-4mm}0&\hspace{-5mm}0 & \hspace{-1mm}{c_k g_{k\!-\!1} r_2 t_2 \over c_{k\!-\!1} g_k} & \hspace{-3mm}0 \\[2mm]
0&0&\hspace{-2mm}0&\hspace{-4mm}0&\hspace{-4mm}0&\hspace{-5mm}0 &0&\hspace{-5mm}{ g_{k\!-\!1}   r_2 t_2 \over c_{k\!-\!1}  g_k}\\[2mm]
0&0&\hspace{-2mm}0&\hspace{-4mm}0&\hspace{-4mm}0&\hspace{-5mm}0&0&\hspace{-6mm}{ g_{k\!-\!1} s_2 r_2  t_2 \over  g_k}\!\! \\[2mm]
0&0&\hspace{-2mm}0&\hspace{-4mm}0&\hspace{-4mm}0&\hspace{-5mm}0&0&\hspace{-3mm}0 
\end{pmatrix}
\label{P30}
\end{equation}
\end{widetext}
The first four columns reproduce the two magnon solution~\eqref{P20}.
The coefficients $c_k$, $f_k$ and $d_k$
are determined by~\eqref{ccoeff}-\eqref{dcoeff}, with the initial conditions~\eqref{icond}.
The new coefficients $h_k$ and $g_k$ are determined by the recursion relations
\begin{eqnarray} 
h_k f_{k\!-\!1} &= & h_{k\!-\!1} f_k \, t_2 r_2^\ast +d_{k\!-\!1}^2 B_k \ ,\\[1mm]
B_k& = & B_{k\!-\!1} \, t_2 s_2^\ast +f_{k\!-\!1}   \ ,\\[1mm]
d_{k\!-\!1} g_k^2 & =& |h_k|^2 + d_k^2 \, g_{k\!-\!1}^2 |t_2|^2  \ ,
\end{eqnarray}
with $g_k$ real and positive.
The initial conditions for these equations are contained in $G_1$. The simplicity of the XX model leads to the efficient decomposition of the unitaries $P_k$ in terms of a single layer of $Fsim$ gates.
It is straightforward to check that the previous two magnon and three magnon matrices verify the ansatz in~\eqref{eqn:ansatz}. The parameters~\eqref{eqn:F} describing the corresponding gates $F_{j,k}$ are
\begin{eqnarray}
a_{M,k}& \hspace{-8mm}={ 1 \over c_k} \ , \hspace{3mm} & \hspace{5mm} b_{M,k}={ c_{k\!-\!1} s_2 \over c_k} \ ,  \label{parI}\\[1mm]
a_{M\!-\!1,k}&=-{ f_k \over c_{k\!-\!1} d_k} \ , \hspace{5mm} &b_{M\!-\!1,k} ={ c_k d_{k\!-\!1} r_2 \over c_{k\!-\!1} d_k} \ , \label{part2}\\[1mm]
a_{M\!-\!2,k}& \!\!\!\!={ h_k \over d_{k\!-\!1} g_k} \ , \hspace{5mm} & b_{M\!-\!2,k}={d_k  g_{k\!-\!1}  t_2 \over d_{k\!-\!1} g_k} \ , \;\;\;\; \label{parF}
\end{eqnarray}
with $a=\cos \theta \, e^{i \alpha}$ and $b=\sin \theta \, e^{i \beta}$.
For $M=2$, only~\eqref{parI} and~\eqref{part2} are relevant.

\section{Ground state of the XX model on \texorpdfstring{$4$}{4} sites}
\label{app:XX_GS}
\begin{figure}[ht!]
    \centering
    \includegraphics[width=\columnwidth]{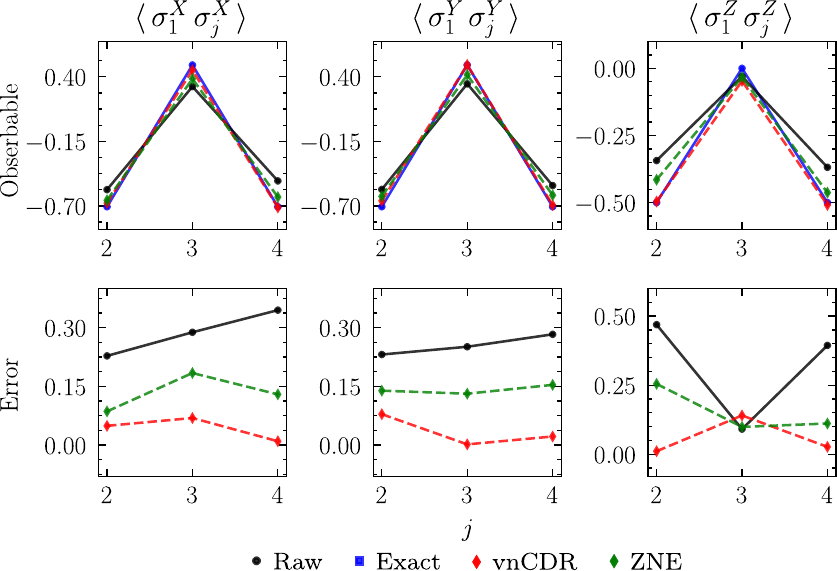}
    \caption{Two-point correlators for the ground state of the XX model on $4$ sites calculated using the quantum computer  \textit{IBM\_Montreal}. The first row shows the correlators, while the second row shows their weighted error, defined in Eq.~\ref{eqn:error}. The raw observables were calculated to have a mean weighted error of $0.29$, ZNE reduced this error to $0.14$ and vnCDR to $0.05$.}
    \label{fig:correlators2mag_XX}
\end{figure}

We implement here the circuit construction for the ground state of the XX model for $4$ sites. The circuit consists of $5$ $F$ gates, its depth is $40$ and involves $10$ CNOTs. Fig.~\ref{fig:correlators2mag_XX} shows the two-point correlators together with the error associated with each mitigation method. We find that the performance of vnCDR is better than ZNE as indicated in the main text.

\section{Example of \texorpdfstring{$\mathscr{R}_{\sigma}$}{Rsigma} with 3 magnons}
\label{app:R_sigma}
The matrix ${\mathscr R}_\sigma$ appearing in~\eqref{eqn:Rsigma} is the essential element for the derivation of the unitary version of the YB equation. 
We present here an example of this  matrix for a permutation involving three magnons.
We assign the following rapidities to the $R$ matrices in the basic cell~\eqref{MM3}, 
\begin{equation}
    R_3=R(\lambda) \ , \;\;\;\; R_2=R(\mu)\ , \;\;\;\;
    R_1=R(\nu) \ ,
\end{equation}
and consider the permutation $\sigma(\lambda,\mu, \nu)=(\nu,\mu,\lambda)$. Using the YB equation~\eqref{eq:YB}, this exchange is achieved by
\begin{equation}
    \vcenter{\hbox{\includegraphics[width=0.85\columnwidth]{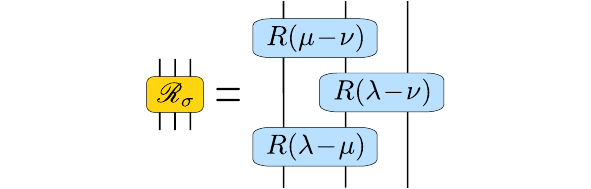}}} \ .
\end{equation}
The construction of ${\mathscr R}_\sigma$
is not unique, since
\begin{equation}
    \vcenter{\hbox{\includegraphics[width=0.85\columnwidth]{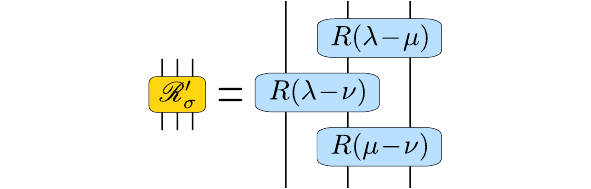}}}
\end{equation}
has the same effect. Consistently, the YB equation guarantees that ${\mathscr R}_\sigma={\mathscr R}'_\sigma$.

\section{Two-magnon exchange matrices}
\label{exchange}
We construct here the exchange matrices $M_{k}$ implementing the unitary version of the YB equation for $M=2$
\begin{equation}
      \vcenter{\hbox{\includegraphics[width=0.85\columnwidth]{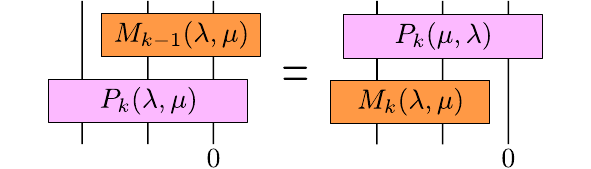}}} \ ,
\end{equation}
where 
\begin{equation}
    M_k(\lambda,\mu)=
    G_k(\mu,\lambda)\,R(\lambda-\mu)\, G_k^{-1}(\lambda,\mu) \ .
\end{equation}
The rapidity $\lambda$ is associated with the matrix $R_2$ in~\eqref{RT2}. Following the notation there, we call its entries  $s_{1,2}$. The rapidity $\mu$ determines the matrix $R_2$, with entries $r_{1,2}$. The difference of rapidities $\lambda -\mu$ defines an $R$ matrix whose entries we denote as $t_{1,2}$.
The YB equation~\eqref{eqn:YB} determines
\begin{equation}
    t_1= {s_1 r_1 \over r_1^2 -r_2^2+ s_2 r_2} \ , \hspace{5mm} t_2= {s_2 -r_2 \over r_1^2 -r_2^2+ s_2 r_2} \ .
\end{equation}
From the expressions in Appendix~\ref{app:XXZ2_model}, we obtain
\begin{equation}
M_{k}=\begin{pmatrix}
1&0&0&0 \\[2mm]
0& {c_k(s_2,r_2) \over c_k(s_2) c_k(r_2)}& {(s_2^\ast -r_2^\ast )\, d_k  \over c_k(s_2) c_k(r_2)} & 0 \\[2mm]
0 & {(r_2 -s_2 ) \, d_k \over c_k(s_2) c_k(r_2)}  & {c_k(r_2,s_2) \over c_k(s_2) c_k(r_2)} & 0 \\[2mm]
0 & 0 & 0 & 1
\end{pmatrix} \ , 
\end{equation}
We have generalized~\eqref{ccoeff} to
\begin{equation}
    c^2_k(s,r) = c_{k\!-\!1}^2(s,r) \,s \, r^\ast +1 \ ,
\end{equation}
subject to the initial condition $c_0(s,r)=1$. Hence $c_k(s_2)=c_k(s_2,s_2)$ coincides with the previously defined coefficients $c_k$, and $c_k(r_2)=c_k(r_2,r_2)$ is assumed real and positive. In deriving the above expression we have used that $d_k$~\eqref{dcoeff} and $e_k$~\eqref{ehorror} are symmetric under the exchange of $s_2$ and $r_2$. Our main result here is the independence of the exchange matrices $M_k$ on the anisotropy.\\

\section{Error mitigation details}
\label{app:mitigation}
For detailed analysis of the various mitigation techniques we refer the reader to Refs.~\cite{zne, cdr, vncdr}. In particular, the implementations here are very similar to those in Ref.~\cite{quench_2021}. We use identity insertions~\cite{He_2020} to scale the noise by a factor of $3$. Therefore, we use noise levels ${1,3}$ in both ZNE and vnCDR. We note that scaling the noise using identity insertions is not optimal, we expect the implementation of ZNE and vnCDR presented here could be improved upon using pulse stretching. 

For the training circuits used in vnCDR we replace half of the non-Clifford gates, selecting them randomly and replacing them probabilistically with a Clifford gate as detailed in Refs.~\cite{cdr,vncdr}. For the  results taken from \textit{IBM\_Montreal} We use $100$ training circuits, and all circuits were run with $32000$ shots. Therefore, there is a shot overhead for the implementations of CDR and vnCDR over ZNE by factors of roughly $25$ and $50$ respectively. For the data obtained from the \textit{IBM\_Mumbai} computer we use $8192$ shots with $48$ training circuits. Therefore, there is a shot overhead for the implementations of CDR and vnCDR over ZNE by factors of roughly $12$ and $25$ in this case. These parameters reflect the greatest possible numbers of shots and different circuits that can be run in one job. We note that we repeated the experiments shown in the main text several times on different devices. We present the results where the noise is most stable and well behaved. In all our runs we found vnCDR performs the best on average.

Overall, we find each mitigation method we explore is successful in mitigating the effect of noise. However, even for these small systems it is apparent further techniques should be combined to remove the effect of noise further. We note that our circuit compilation strategies can most likely be improved to reduce circuit depth. Furthermore, a combination of dynamical decoupling~\cite{1998_DD} and various other error mitigation techniques has been shown to produce accurate observables of interest~\cite{kim2021scalable}. It would be interesting to apply a similar approach in the context of ABCs.
\end{document}